\documentclass[useAMS,usenatbib]{mn2e_mod}
\usepackage{natbib}
\usepackage{graphicx}
\usepackage{amsmath}
\usepackage{amssymb}
\usepackage{deluxetable}
\usepackage{longtable}

\newcommand{\cloudy}{{\sc cloudy}}

\let\ACMmaketitle=\maketitle
\renewcommand{\maketitle}{\begingroup\let\footnote=\thanks \ACMmaketitle\endgroup}

\title [AGN feedback with KCWI]{Direct evidence of AGN-feedback: a post starburst galaxy stripped of its gas by AGN-driven winds. \footnote{The data presented herein were obtained at the W.M. Keck Observatory, which is operated as a scientific partnership among the California Institute of Technology, the University of California and the National Aeronautics and Space Administration. The Observatory was made possible by the generous financial support of the W.M. Keck Foundation.}}

\author[Baron et al.]
{Dalya Baron$^{1}$\thanks{dalyabaron@mail.tau.ac.il},
Hagai Netzer$^{1}$,
J. Xavier Prochaska$^{2}$,
Zheng Cai$^{2}$,
Sebastiano Cantalupo$^{3}$,
\newauthor 
D. Christopher Martin$^{4}$,
Mateusz Matuszewski$^{4}$,
Anna M. Moore$^{5}$,
\newauthor 
Patrick Morrissey$^{4}$ \&
James D. Neill$^{4}$
\\
\\
$^{1}$School of Physics and Astronomy, Tel-Aviv University, Tel Aviv 69978, Israel.\\
$^{2}$Department of Astronomy and Astrophysics, UCO/Lick Observatory, University of California, 1156 High Street, Santa Cruz, CA 95064, USA.\\
$^{3}$Department of Physics, ETH Zurich, Wolfgang-Pauli-Strasse 27, 8093 Zurich, Switzerland.\\
$^{4}$California Institute of Technology, 1216 East California Boulevard, Pasadena, CA 91125, USA.\\
$^{5}$Research School of Astronomy and Astrophysics, Australian National University, Canberra, ACT 2611, Australia.
}

\begin{document}

\maketitle

\label{firstpage}
\begin{abstract}

Post starburst E+A galaxies show indications of a powerful starburst that was quenched abruptly. Their disturbed, bulge-dominated morphologies suggest that they are merger remnants. The more massive E+A galaxies are suggested to be quenched by AGN feedback, yet little is known about AGN-driven winds in this short-lived phase. We present spatially-resolved IFU spectroscopy by the Keck Cosmic Web Imager of SDSS J003443.68+251020.9, at z=0.118. The system consists of two galaxies, the larger of which is a post starburst E+A galaxy hosting an AGN. Our modelling suggests a 400 Myrs starburst, with a peak star formation rate of 120 $\mathrm{M_{\odot}}$/yr. The observations reveal stationary and outflowing gas, photoionized by the central AGN. We detect gas outflows to a distance of 17 kpc from the central galaxy, far beyond the region of the stars ($\sim$ 3 kpc), inside a conic structure with an opening angle of 70 degrees. We construct self-consistent photoionization and dynamical models for the different gas components and show that the gas outside the galaxy forms a continuous flow, with a mass outflow rate of about 24 $\mathrm{M_{\odot}}$/yr. The gas mass in the flow, roughly $10^{9}\, \mathrm{M_{\odot}}$, is larger than the total gas mass within the galaxy, some of which is outflowing too. The continuity of the flow puts a lower limit of 60 Myrs on the duration of the AGN feedback. Such AGN are capable of removing, in a single episode, most of the gas from their host galaxies and expelling enriched material into the surrounding CGM.

\end{abstract}


\vspace{1cm}
\section{Introduction}\label{s:intro}

The discovery of a strong correlation between the masses of super massive black holes (SMBHs) and the velocity dispersions and bulge mass of their host galaxies \citep{gebhardt00a, ferrarese00, tremaine02, gultekin09} led to various suggestions about the connection between SMBH growth and stellar mass growth in galaxies, in particular the quenching of star formation (SF) due to BH activity. Such feedback is now a key ingredient in galaxy formation theories \citep{silk98, springel05a, springel05b, hopkins06}. Several such models include a scenario in which an active galactic nucleus (AGN) drives galactic-scale winds that expel gas from its host galaxy, shuts down additional gas accretion onto the SMBH, terminates the star formation in the galaxy, and enriches its circumgalactic medium (CGM) with metals \citep{silk98, fabian99, benson03, king03, dimatteo05, hopkins06, gaspari11}.

AGN-driven winds are ubiquitous and span a large range of host galaxy properties. They are observed in systems at different evolutionary stages, with a large range of star formation rates (SFRs): from ultra luminous infrared galaxies (ULIRGs; \citealt{rupke05, rupke13, spoon13, veilleux13, zaurin13, sun14, burillo15, fiore17}), which show powerful starbursts, to quenched elliptical galaxies (e.g. \citealt{cheung16}). They are detected on different physical scales, from the vicinity of the SMBH (few gravitational radii; \citealt{blustin03, reeves03, tombesi10}) to galactic scales ($\sim$1-10 kpc; e.g., \citealt{greene11, cano12, liu13a, liu13b, harrison14, rupke17}). They are traced via different gas-phase indicators related to different ionization states - from high velocity X-ray and UV absorption lines (e.g. \citealt{blustin03, reeves03, tombesi10, arav13}), through ionized emission lines \citep{greene11, cano12, harrison14, zakamska16, rupke17}, to atomic and molecular emission and absorption lines \citep{feruglio10, veilleux13, cicone14, burillo15, rupke17}. While outflows are detected in both type I and type II AGN \citep{heckman81, feldman82, heckman84, greene05, nesvadba06, moe09, rosario10, greene11, cano12, arav13, mullaney13, zakamska16, fiore17}, it is still unclear whether such winds are capable of expelling a large enough mass of gas from their host galaxies to significantly affect their evolution. Furthermore, there are very few systems in which this scenario was observed in action, i.e. systems in which most of the gas was recently driven out of the galaxy (see examples of possible candidates in \citealt{greene11, liu13b, sun18}). At high redshift of $z\approx2$, \citet{cai17} identified a $\gtrsim30$ kpc extended CIV and He II emission in the radio-quiet MAMMOTH-1 enormous nebula (ELAN), which they interpret as a direct evidence of an AGN outflow at high-redshift.

A major uncertainty concerning galactic-scale feedback is the relative contribution of AGN-driven versus supernovae-driven winds, where the latter is directly related to SF activity in the galaxy. The current largest samples of observed outflows come from optical surveys, such as the SDSS (\citealt{york00}), where blueshifted and asymmetric emission lines are used to select systems with winds (see e.g. \citealt{mullaney13}). In most of these samples, only the [OIII]$\lambda$5007\AA\, emission line shows a blueshifted profile, which is not enough to determine the ionization source of the winds (SF, AGN, or shocks). Studies either focus on type I AGN, where the accretion disk (AD) is detected directly, or use narrow emission line diagnostics (see e.g. \citealt{kewley06}) to select systems in which there is an obscured type II AGN. Although the AGN is the main ionizing source of the stationary (and sometimes the outflowing) gas, in all such sources, it is unclear whether it is also the main driver of the observed winds. This is due to the fact that most systems showing AGN activity also show significant SF activity, with some correlation between the AGN bolometric luminosity and the SF luminosity (e.g. \citealt{netzer09}). Many systems with more powerful AGN, that are capable of producing stronger AGN-driven winds, also undergo powerful SF episodes, capable of driving stronger supernovae-driven winds. 

A similar uncertainty is associated with pure SF galaxies selected by line diagnostic diagrams (see a comprehensive study by \citealt{cicone16}). These systems do not show evidence of current AGN activity, but due to the relatively short AGN duty-cycle it is not clear whether the AGN was active in the recent past and might have contributed to the currently observed winds. Thus, the timescale of the observed winds, and the relation to AGN and/or SF activity, are crucial for the understanding of the various types of feedback. As explained below, we suggest that detailed IFU observations of relatively nearby post-starburst galaxies can help to answer such questions.

Post starburst E+A galaxies (also called H$\delta$-strong and K+A galaxies) offer an advantage over other galaxy samples in dealing with these uncertainties \citep{wild10}. These systems show prominent Balmer absorption lines in their optical spectra, and no contribution from O and B-type stars \citep{dressler99, poggianti99, goto04, dressler04}. Their spectrum indicates a recent starburst that was quenched abruptly, with a narrow stellar age distribution, completely dominated by intermediate-age stars (typically A-type stars). The estimated SFRs during the burst range from 50 to 300 $\mathrm{M_{\odot}/yr}$ \citep{poggianti00, kaviraj07}, and the mass fractions forming in the burst are high, 30\%--80\% of the total stellar mass \citep{liu96, bressan01, norton01, yang04, kaviraj07}. Many of these systems show bulge-dominated morphologies, with tidal features or close companions, which suggests a late-stage merger \citep{canalizo00, yang04, goto04, cales11}. Due to the short lifetime of A-type stars, this evolutionary stage must be very short, making such systems a small fraction of the total galaxy population \citep{goto03, goto07, wild09, yesuf14, alatalo16a}.

Various studies suggest that post starburst E+A galaxies are the evolutionary link between gas rich major mergers (ULRIGs) and quiescent, early-type, galaxies \citep{yang04, yang06, kaviraj07, wild09, cales11, cales13, yesuf14, cales15, french15, alatalo16a, alatalo16b, wild16, baron17b}. According to this scenario, a gas-rich major merger triggers a powerful starburst, and gas is funneled to the vicinity of the SMBH, triggering an AGN. Soon after, the AGN launches nuclear winds which sweep-up the gas in the galaxy, shutting down the current starburst abruptly, and removing the gas from the host galaxy \citep{kaviraj07, cales13, cales15}. The system is observed as a post starburst E+A galaxy, which soon becomes a quiescent elliptical galaxy. \citet{kaviraj07} performed the first comprehensive observational and theoretical study of the quenching process that terminates the starburst in E+A galaxies. They found a bimodal behaviour, where in galaxies less massive than $\mathrm{M = 10^{10}\,M_{\odot}}$, SN-driven feedback is the main quenching mechanism, while galaxies above this mass are mainly quenched by an AGN (see also \citealt{li18}). They suggested, based on calculated star formation histories (SFHs), that the latter were indeed ULIRGs. According to this scenario, one may expect to detect galactic-scale AGN-driven winds in these systems.

For E+A galaxies with an active BH and ongoing observed winds, it is clear that the observed outflows are driven solely by the AGN, since the starburst is fully quenched. This allows the study of wind properties and dynamics in the context of pure AGN feedback. Furthermore, since the stellar age throughout the entire galaxy is dominated by a single, short, starburst episode, it allows a comparison between winds observed in different systems as a function of time since the onset (or termination) of the burst \citep{wild10}. The comparison of several similar systems of this type can serve as a well-defined timeline, that provide the necessary missing details for AGN-feedback in this phase. 

In recent years, studies using optical and NIR integral field units (IFUs) have provided detailed information about spatially-resolved outflows that are traced by ionized gas \citep{greene11, cano12, liu13a, liu13b, harrison14, rupke17}. However, observations of high redshift systems of this type cannot resolve scales below several kpc and in low redshift systems, there is a confusion about the source of the observed winds with contributions from both AGN and stellar radiation fields (e.g. \citealt{harrison14}). It is essential to improve the observations and modeling of the nearby systems and also focus on a well-defined sample of objects (starburst galaxies, post-starburst galaxies, mergers, etc.), since feedback can take different forms, depending on the morphology of the system and its (usually undetermined) age. This work, which is the second in a series of papers on feedback in E+A galaxies, provides many of the missing details for one such system, using optical IFU observations.

We have recently detected the first evidence of an AGN-driven outflow, traced by ionized gas, in a post starburst E+A galaxy (\citealt{baron17b}; see also \citealt{tremonti07}, \citealt{tripp11}, and \citealt{yesuf17b}). SDSS J132401.63+454620.6 was discovered as an outlier by the anomaly detection algorithm of \citet{baron17}, and our ESI/Keck spectroscopy revealed a post starburst system with powerful ionized outflows. We measured the mass outflow rate in this system to be in the range 4--120$\mathrm{M_{\odot}/yr}$, similar to outflows observed in ULIRGs. Since then, we have constructed a sample of such galaxies, with fully-quenched starbursts and powerful AGN-driven winds. In this work we present spatially resolved spectroscopy, obtained by the Keck Cosmic Web Imager (KCWI), of a second E+A galaxy at z=0.118, SDSS J003443.68+251020.9. Our observations reveal ionized winds that extend far beyond the host galaxy, with most of the gas mass already detached from the main stellar mass. We describe the observations in section \ref{s:obs}, and discuss the general observed properties of the system in section \ref{s:system_properties}. We then construct a photoionization and dynamical model for the system in section \ref{s:photoionization_model}, and measure masses and mass outflow rates in section \ref{s:masses}. We discuss our finding in section \ref{s:disc} and conclude in section \ref{s:concs}. Throughout this paper we assume a cosmology with $\Omega_{\mathrm{M}}=0.3$, $\Omega_{\Lambda}=0.7$, and $h=0.7$, thus 1'' corresponds to 2.1 kpc for the system in question.

\section{Observations}\label{s:obs}

\subsection{1D spectroscopy from SDSS}

SDSS J003443.68+251020.9 was observed as part of the general SDSS survey \citep{york00}. The spectrum was obtained using the BOSS spectrograph with a 2'' fiber, which covers a wavelength range of 3800\AA--9200\AA, with a resolving power going from 1560 at 3700\AA\, to 2650 at 9000\AA, resulting in a spectral resolution of 190 km/sec and 110 km/sec respectively. The publicly available spectrum of the galaxy is combined from three 15m sub-exposures, and with signal to noise ratios (SNR) ranging from 10 to 25. We did not find spectral differences between the 3 sub-exposures. The image of the galaxy was taken in the SDSS broad band filters with a typical seeing of 1''--1.2''.

\subsection{Spatially-resolved spectroscopy with KCWI}

The Keck Cosmic Web Imager (KCWI) is a new facility instrument for the Keck II telescope at the W. M. Keck Observatory (WMKO; \citealt{martin10, morrissey12}; Morrissey et al. in prep.). KCWI is a bench-mounted spectrograph installed at the Keck II right Nasmyth focal station, providing integral field spectroscopy over a seeing-limited field of up to 20''$\times$33'' in extent, with high efficiency and spectral resolution in the range 1000--20000. KCWI will provide full wavelength coverage (0.35 to 1.05 $\mu m$) using optimized blue and red channels, with the blue channel (0.35 to 0.55 $\mu m$) currently in operation. 

We observed the system on the night of 2017 October 20. Conditions were clear with seeing of approximately 1.2". Observations were carried out with the medium slicer, which gives a 16.5''x20.4'' field of view, with a slice width of 0.69''. We used the BM grating configuration, tilted to provide wavelength coverage of 4600–-5700\AA, with a spectral resolution of 0.28\AA. Three exposures of 600s provided a combined binned data cube, with a median SNR per spatial pixel (spaxel) between 2 and 12 in the regions where a source is detected. The average surface brightness limit of the combined cube is about $1.5 \times 10^{-18}\,\mathrm{erg\,sec^{-1}cm^{-2}\AA^{-1}\,arcsec^{-2}}$.

The standard KCWI pipeline\footnote{KCWI pipeline: https://github.com/kcwidev/kderp/blob/master. A forthcoming paper will describe the KCWI pipeline in detail (Neill et al. in prep).} was used to reduce the data (also see brief description in Cai et al. in prep.). For each datacube, the bias was subtracted, the pixel-to-pixel variation was corrected, and the cosmic-rays were removed. We corrected the geometric distortion using the continuum bar images. The arc images were used to calibrate the wavelength solution. The slice-to-slice variance was further removed using the twilight flats. We use the BD+28D4211 taken at the beginning of the night to conduct the flux calibration. 

In order to subtract sky emission lines, we choose three different off-source regions and bin their spectra. We subtract the binned spectrum from all the spaxels in the cube. Once the sky emission is subtracted, an integration over 2'' around the primary galaxy gives a spectrum that is similar to that by the SDSS.

\section{General observed properties}\label{s:system_properties}
\begin{figure}
\includegraphics[width=3.25in]{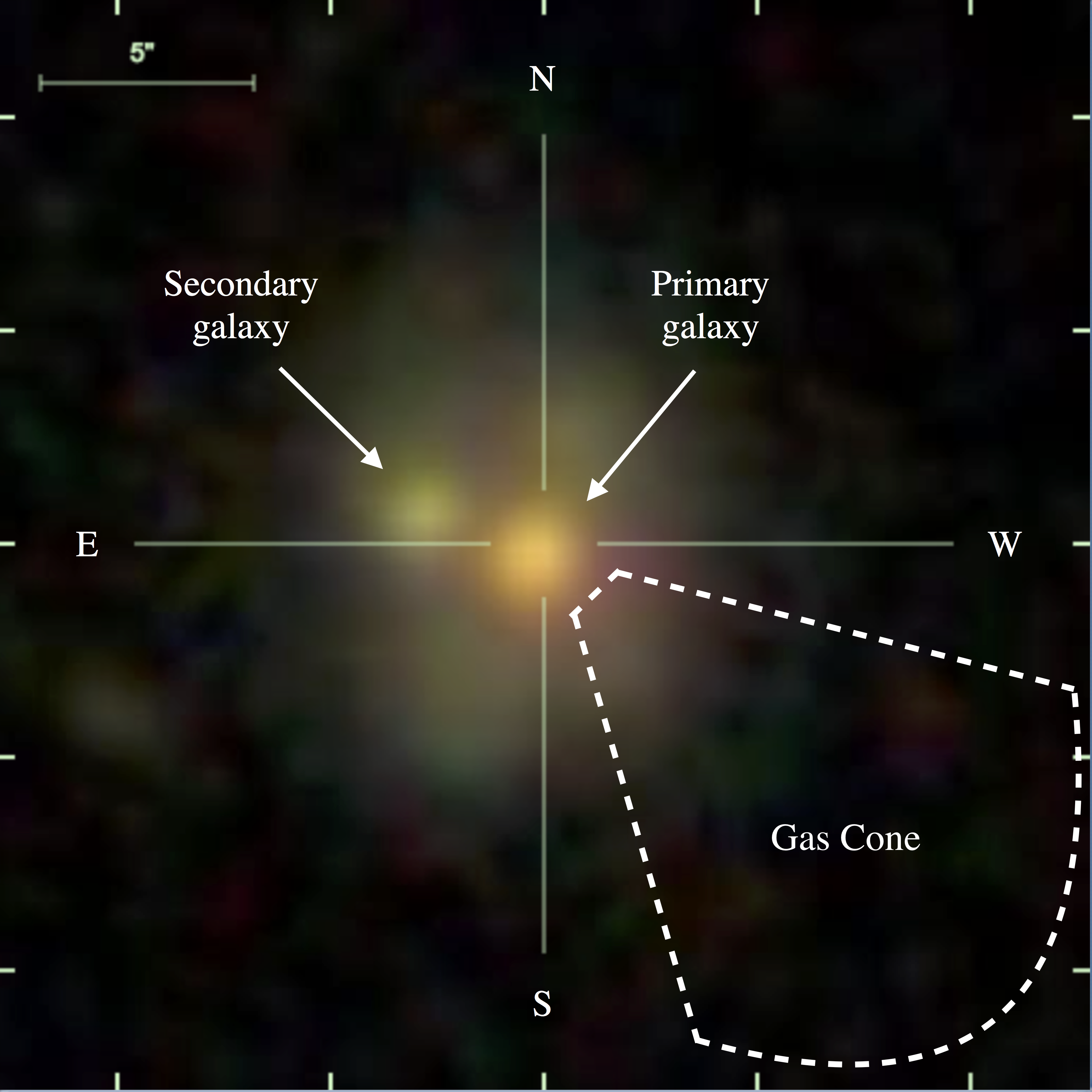}
\caption{The $gri$ color-composite image of SDSS J003443.68+251020.9. The image shows two different galaxies in the field, which we call the primary and secondary galaxies. An SDSS spectrum is available only for the primary galaxy. The white dashed lines mark the region in which we detect ionized gas emission.}\label{f:galaxy_image}
\end{figure}

In figure \ref{f:galaxy_image} we show the SDSS $gri$ color-composite image of the system. The system is composed of two galaxies, which we refer to as the primary and the companion. An SDSS spectrum is available only for the primary galaxy. We study the stellar properties of the primary in section \ref{s:galaxy_properties}, and provide details about the companion in section \ref{s:companion}. The properties of the AGN and the gas in the system are described in section \ref{s:agn_properties}. Throughout the section, we use both SDSS and KCWI data. While KCWI data is superior in terms of its spectral resolution, spatial resolution, and SNR, it covers a limited wavelength range compared to the SDSS spectrum. The SDSS spectrum is an integration through a 2'' fiber around the center of the primary galaxy. We refer to it throughout the section as the "global spectrum". 

\subsection{The primary galaxy}\label{s:galaxy_properties}
The SDSS color-composite image (figure \ref{f:galaxy_image}) shows that the primary galaxy has a bulge-dominated morphology, with a nearby companion, and additional diffuse emission around the two systems. Thus, we most likely observe a galaxy interaction. We fit a Sersic profile to the primary images in the $g$ and $r$ bands and find Sersic indices of 4.9 and 4.1 respectively. These indices are higher than what is typically found in elliptical (Sersic index of 3--4) and spiral (Sersic index of 1) galaxies, which means that the primary galaxy is more concentrated than typical ellipticals and spirals. 

\begin{figure*}
\includegraphics[width=0.9\textwidth]{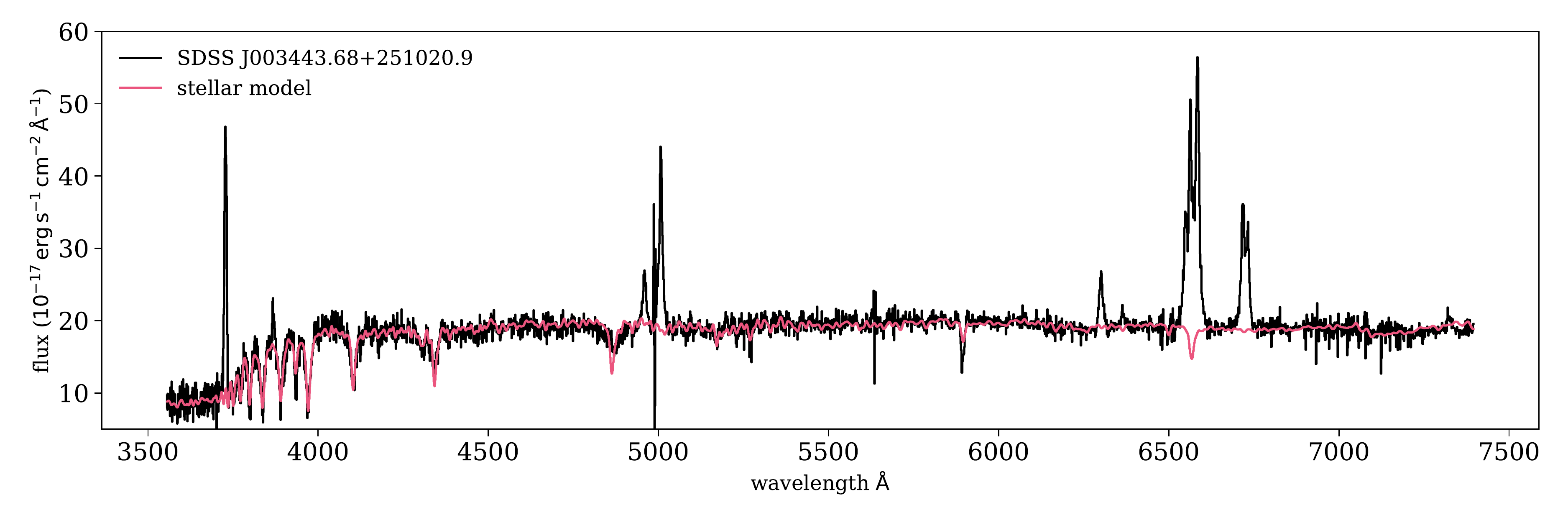}
\caption{The SDSS 1D spectrum of SDSS J003443.68+251020.9 (black), and the best population synthesis model (pink) using pPXF.}\label{f:stellar_pop}
\end{figure*}

In figure \ref{f:stellar_pop} we show the global spectrum of the primary galaxy. The spectrum is dominated by A-type stars and shows strong Balmer absorption lines, the clear signature of post starburst E+A galaxies. The equivalent width (EW) of the H$\delta$ absorption line is 7.8\AA, which is above the 5\AA\, threshold typically used to select E+A galaxies (see e.g., \citealt{goto07, alatalo16a}). We fitted a stellar population synthesis model to the spectrum using the Penalized Pixel-Fitting stellar kinematics extraction code (pPXF; \citealt{cappellari12}), which is a public code for extracting the stellar kinematics and stellar population from absorption-line spectra of galaxies \citep{cappellari04}. It uses the MILES library, which contains single stellar population (SSP) synthesis models and covers the full range of the optical spectrum with a resolution of full width at half-maximum (FWHM) of 2.3\AA\, \citep{vazdekis10}. The output of the code includes the relative weight of stars with different ages, the stellar velocity dispersion, the dust reddening towards the stars (assuming a \citealt{calzetti00} extinction law), and the best-fitting stellar model, which we mark with pink in figure \ref{f:stellar_pop}.

The best-fit model shows a stellar age distribution which consists of two SF episodes, with a total mass-weighted stellar age of 0.28 Gyr. According to the model, the recent episode, which dominates the spectrum of the galaxy, started 400 Myrs ago and ended 200 Myrs ago, with peak SFR of 120 $\mathrm{M_{\odot}/yr}$. There is no contribution from O and B-type stars to the model, which means that there is no ongoing SF in the galaxy. The older SF episode is poorly constrained due to its weak contribution to the spectrum. The age of the older episode is centered around 5 Gyr, with a width of 2 Gyr. However, we find reasonable fits even when forcing the code to use templates with different ages within the range 3 -- 10 Gyrs. The dust reddening towards the stars is $\mathrm{E}(B-V) = 0.45$ mag, within the range observed in star forming galaxies \citep{kauff03b}. The stellar velocity dispersion is 170 km/sec, which is well-resolved due to the relatively broad absorption lines in A stars. We use the best stellar population synthesis model to measure the stellar mass, which is $\mathrm{log\,M/M_{\odot}} = 10.8$. We use the KCWI data to show below that the stellar continuum of the primary galaxy is detected to 3 kpc. Therefore, the steepness of the light profile suggests that the fiber-based stellar mass measurement, which covers 4.3 kpc, is a good approximation of the total stellar mass of the system.

\subsection{The companion galaxy}\label{s:companion}
\begin{figure}
\includegraphics[width=3.25in]{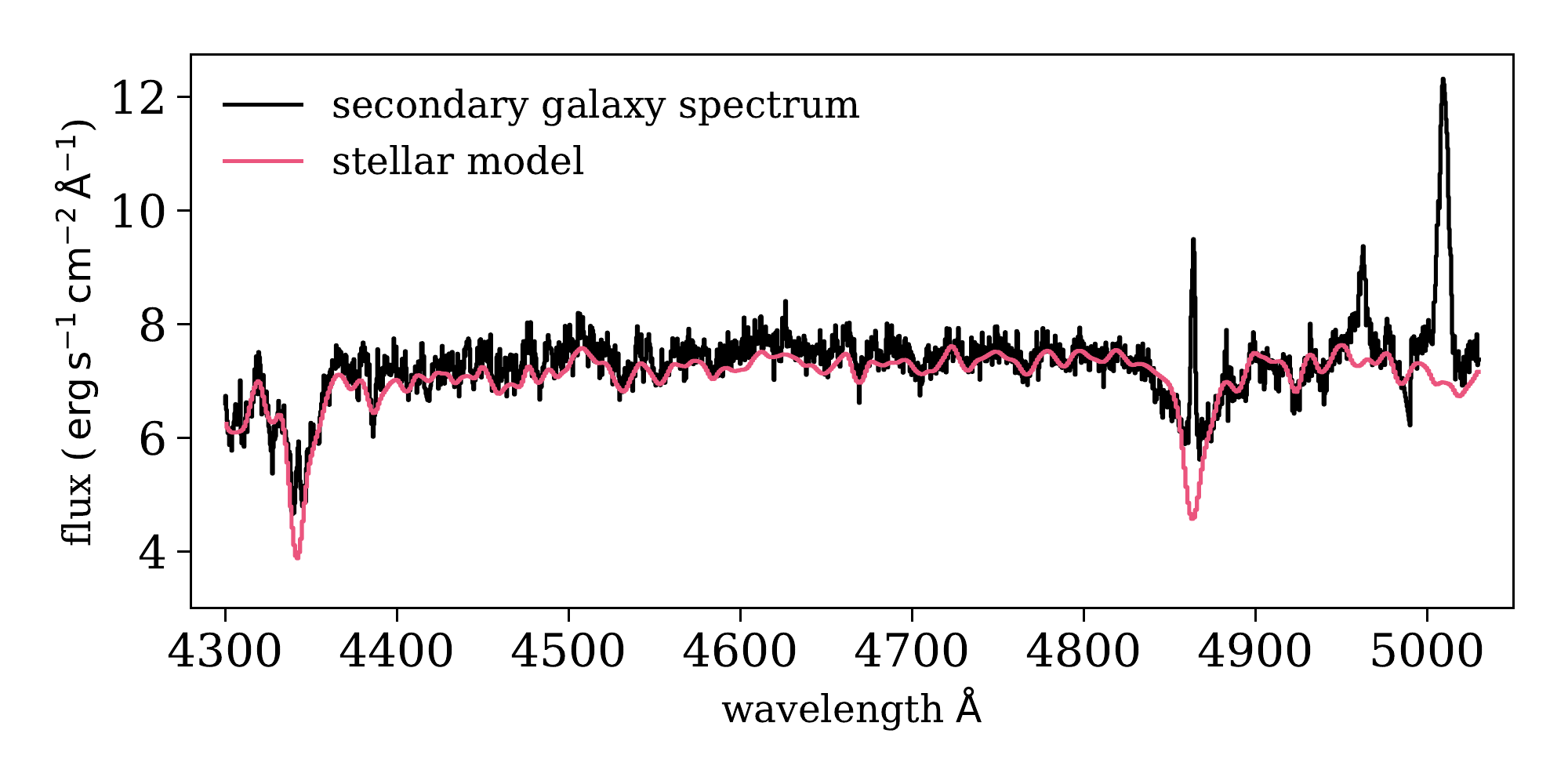}
\caption{The summed KCWI spectrum of the companion galaxy over a region of 2''$\times$2'' (black), and the best population synthesis model (pink) using pPXF. Although the wavelength range is limited, the stellar spectrum and its best fit are similar to what we find for the primary galaxy.}\label{f:stellar_pop_secondary}
\end{figure}

As noted above, there is no SDSS spectrum for the companion galaxy. In order to study its properties we use the KCWI data, and sum all the spaxels within a 2''x2'' region around its center. The summed spectrum is shown in figure \ref{f:stellar_pop_secondary}. We use pPXF to fit stellar population synthesis model to the galaxy, and mark the best fit in figure \ref{f:stellar_pop_secondary} with pink. The best-fit model shows a good agreement with the summed spectrum. However, a large combination of parameters give practically the same stellar continuum, and given the limited wavelength range we cannot robustly determine the age of the recent burst, its duration, or the dust reddening. The strength of the H$\gamma$ absorption line (EW $\sim$ 10\AA) hints that this could also be a post starburst system. This can only be confirmed by observations with a larger wavelength coverage.

One can also see in figure \ref{f:stellar_pop_secondary} that the H$\beta$ emission line is almost as strong as [OIII]. Therefore, we can rule out Seyfert-type ionization for the gas in the secondary galaxy. Based on the available data, we cannot distinguish between SF or LINER-like radiation field. We further note that the two galaxies are unresolved in the WISE images, therefore we cannot use W1, W2, W3, and W4 as further diagnostics for the source of ionization in this galaxy. 

\subsection{Active black hole and AGN photoionized gas}\label{s:agn_properties}
The spectrum of the primary galaxy is very similar to that studied by \citet{baron17b}, where we found a post starburst E+A galaxy with AGN. The emission line spectrum of the primary galaxy is typical of type-II (obscured) AGN. The ionized gas can be roughly divided into three components: (1) the narrow line region within the primary galaxy, (2) a broad kinematic component within the primary galaxy, and (3) gas that resides outside the primary galaxy to distances of 17 kpc. We study the gas within the galaxy in section \ref{s:narrow_and_broad_comps}. We then use the spatially-resolved KCWI data to study the gas outside the galaxy in section \ref{s:gas_properties}.

\subsubsection{The NLR and broad outflowing central component}\label{s:narrow_and_broad_comps}

We subtract the best-fitting stellar model from the global spectrum and obtain the emission-line spectrum of the central gas component. We show six parts of this spectrum in appendix \ref{a:app} (figure \ref{f:1d_spec_line_fit}), where we plot [OIII]~$\lambda \lambda$ 4959,5007\AA, $\mathrm{H\alpha}$~$\lambda$ 6563\AA\, and [NII]~$\lambda \lambda$ 6548,6584\AA, [OII]~$\lambda \lambda$ 3725,3727\AA\, and $\mathrm{H\beta}$~$\lambda$ 4861\AA, and [OI]~$\lambda \lambda$ 6300,6363\AA\, and [SII]~$\lambda \lambda$ 6717,6731\AA\, (hereafter $\mathrm{[OIII]}$, $\mathrm{H\alpha}$, $\mathrm{[NII]}$, $\mathrm{[OII]}$, $\mathrm{H\beta}$, $\mathrm{[OI]}$, and $\mathrm{[SII]}$). The Balmer lines and the forbidden line profiles show narrow as well as broad components. We also detect [NeIII]~$\lambda$ 3869\AA\, emission, which we do not show in figure \ref{f:1d_spec_line_fit}.

The appendix to this paper shows and explain the procedure used to fit the various line profiles. Here we only refer to the line diagnostic diagrams based on the fits. The new high spatial and spectral resolution observations presented in section \ref{s:gas_properties} are far superior to the SDSS spectrum and hence we defer the more detailed description of the line profile in the different locations to that section.

\begin{figure*}
\includegraphics[width=0.9\textwidth]{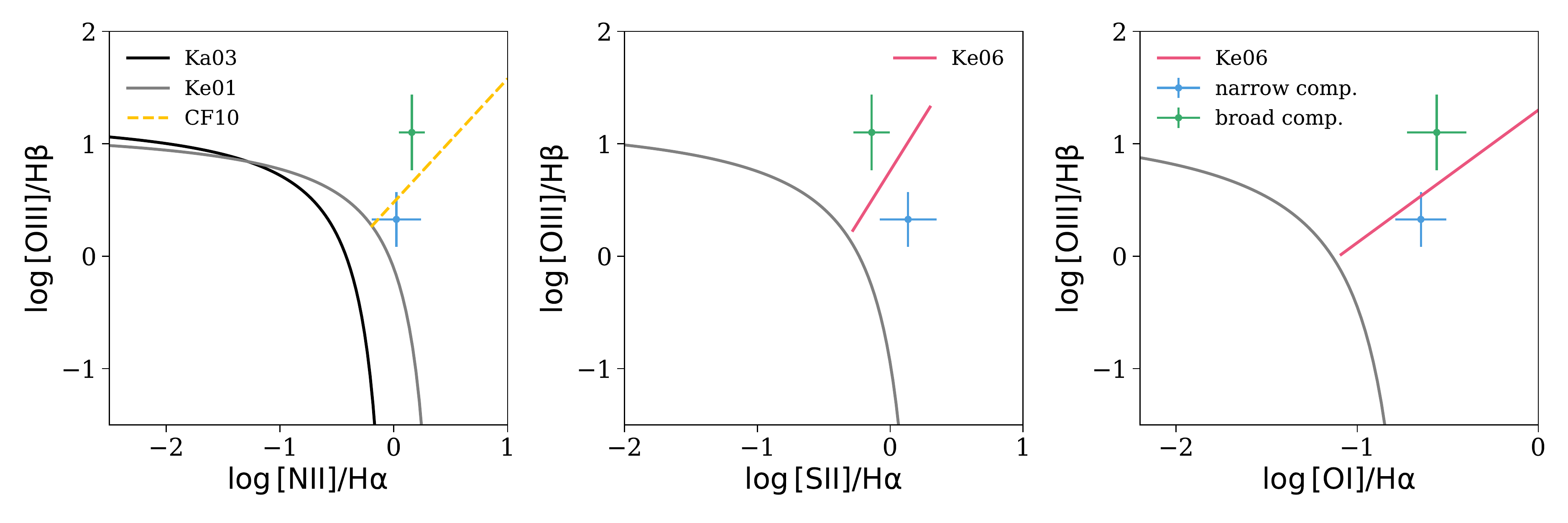}
\caption{Classification diagrams for the narrow (blue) and broad (green) emission lines, as measured from the 2'' SDSS fiber. The left panel shows $\mathrm{log\,[OIII]/H\beta}$ versus $\mathrm{log\,[NII]/H\alpha}$, where we mark the extreme starburst line by Ke01 with black, the composite line by Ka03 with grey, and the LINER-Seyfert separation by CF10 with yellow. The middle and right panels show $\mathrm{log\,[OIII]/H\beta}$ versus $\mathrm{log\,[SII]/H\alpha}$ and $\mathrm{log\,[OI]/H\alpha}$ respectively. We mark the separation between LINER and Seyfert line (Ke06) by a pink line. In all diagrams, the narrow component is consistent with LINER and the broad component is consistent with Seyfert-like emission. }\label{f:1NLR_BPT_diag}
\end{figure*}

In figure \ref{f:1NLR_BPT_diag} we show the narrow (blue) and the broad (green) emission components on line-diagnostic diagrams \citep{baldwin81, veilleux87}. The left panel shows $\mathrm{log\,[OIII]/H\beta}$ versus $\mathrm{log\,[NII]/H\alpha}$. We show three separating criteria, the first is a theoretical upper limit which separates starbursts and AGN-dominated galaxies \citep[Ke01; black line]{kewley01}. The second is a modified criterion which includes composite galaxies showing contributions from both SF and AGN \citep[Ka03; grey line]{kauff03a}. We use the criterion by \citet{cidfernandes10} to separate between LINERs and Seyferts (CF10; yellow). The middle panel and the right panel show $\mathrm{log\,[OIII]/H\beta}$ versus $\mathrm{log\,[SII]/H\alpha}$ and $\mathrm{log\,[OI]/H\alpha}$ respectively. We also mark the \citet{kewley06} separating line between LINER and Seyfert galaxies (Ke06; pink line). One can see that in all three diagrams both the narrow and the broad components are classified as AGN-dominated. The narrow emission lines are consistent with a LINER-type, low ionization, spectrum while the broad lines are consistent with a Seyfert-like, high ionization, spectrum. We note that all these properties are similar to those we observed in SDSS J132401.63+454620.6, the first post starburst E+A galaxy analyzed using such methods (\citealt{baron17b}).

We use the measured H$\alpha$/H$\beta$ flux ratios to derive the global dust reddening towards the two kinematic components. Assuming case-B recombination, a gas temperature of $10^4$ K \citep{osterbrock06}, a dusty screen, and the \citet{cardelli89} extinction law, the colour excess is given by:
\begin{equation}\label{eq:1}
	{\mathrm{E}(B-V) = \mathrm{2.33 log\, \Bigg[ \frac{(H\alpha/H\beta)_{obs}}{2.85} \Bigg] }}
\end{equation}
where $\mathrm{(H\alpha/H\beta)_{obs}}$ is the observed line ratio. Using this relation, we find $\mathrm{E}(B-V) = 0.5$ mag for the narrow lines and $\mathrm{E}(B-V) = 1.7$ mag for the broad lines. The values differ by less than 10\% when using SMC or LMC extinction curves (e.g., \citealt{baron16}). The dust reddening of the narrow component is similar to the reddening derived for the stars, although the two are derived under different geometric assumptions (the stellar reddening is derived with the \citealt{calzetti00} law, which is more appropriate for stars). It is also consistent with the dust reddening typically measured for NLR gas in AGN (see, e.g., \citealt{kewley06}). The dust reddening of the broad component is significantly higher, similar to what we found for SDSS J132401.63+454620.6 in \citet{baron17b}. In section \ref{s:photo_gas_cone} below we look into the emission line reddening in more detail, considering also the case where the dust is mixed with the ionized gas.

\subsubsection{Outflowing extended gas}\label{s:gas_properties}

The KCWI observations cover the rest-frame wavelength range 4200--5120 \AA, allowing us to measure the following emission lines: H$\gamma$, HeII~$\lambda$ 4686\AA, H$\beta$, and the two [OIII] lines. The continuum emission, which is due to the stars in the primary and the secondary galaxies, is similar to the SDSS color-composite shown in figure \ref{f:galaxy_image}. Since we are interested in the gas emission throughout the field, we proceed to remove the stellar continuum emission from each individual spaxel. We use pPXF to fit the stellar continuum in each spaxel. Although the wavelength range is limited, manual inspection of the best-fitting templates show good fits to the observed spectra, and similar templates for different spaxels. The latter is partially due to the PSF broadening. We subtract the best-fitting template from each spectrum to obtain the pure emission line spectrum for each spaxel. 

\begin{figure*}
\includegraphics[width=0.9\textwidth]{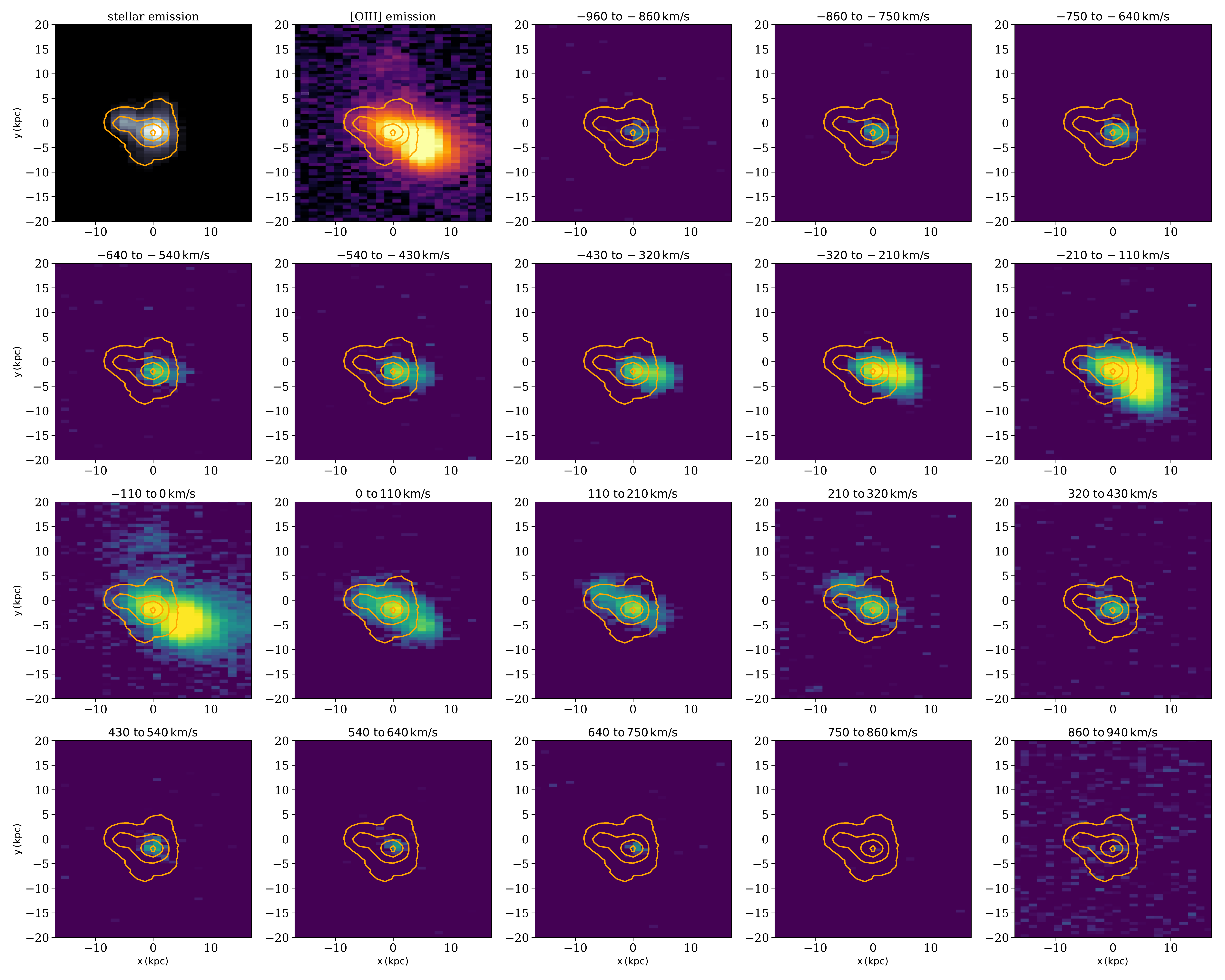}
\caption{First panel: stellar continuum emission with logarithmic color coding in the range $10^{-18}$--$10^{-17}$ $\mathrm{erg\,sec^{-1} cm^{-2}}$ per pixel. Second panel: integrated [OIII] flux with logarithmic color-coding in the range $3 \times 10^{-19}$--$10^{-17}$ $\mathrm{erg\, sec^{-1} cm^{-2}}$ per pixel, where the orange contours represent the stellar continuum level and the largest contour represents the regions in which the continuum level is consistent with pure noise. Next panels: $\mathrm{[OIII]}$ channel maps. Each panel represents a different velocity channel, compared to the systematic velocity of the stars in the primary galaxy, where we sum the $\mathrm{[OIII]}$ flux in bins of about 100 km/sec per channel. The color coding is logarithmic and is kept to a single dynamical range for all the different channels, where yellow is high flux and purple is low flux. The contours trace both the primary galaxy (located at (0, 0) kpc and [OIII] systemic velocity of zero) and the secondary galaxy (located at (-5, 0) kpc and [OIII] systemic velocity of 100 km/sec). The [OIII] emission extends to a distance of 17 kpc from the central galaxy, far beyond the stars in the galaxy ($\sim$3 kpc).}\label{f:oiii_velocity_map}
\end{figure*}

The top two panels of figure \ref{f:oiii_velocity_map} show the stars and the [OIII] emission in the entire system. Stellar light is detected up to a distance of about 3 kpc from the center of the primary galaxy while gas emission extends, in a conic-shaped structure in the South-West direction, up to a projected distance of about 17 kpc. The other panels of figure 5 show the [OIII] flux for different velocity channels, compared to the systematic velocity of the stars in the primary galaxy. Each velocity channel covers about 100 km/sec and we sum the observed [OIII] in this range. We use a logarithmic color coding and keep the same dynamical range for all the different panels, where yellow represents high flux and purple represents low flux. We overplot with orange contours the stellar continuum emission, where the largest orange contour represents the region which is consistent with non-detection. The primary galaxy shows [OIII] emission with a systematic velocity of close to zero and the secondary galaxy (located at -5,0 kpc) shows [OIII] with a systematic velocity of approximately +200 km/sec. Beyond the two galaxies, one can see [OIII] emission with velocities that range between -900 km/sec to +600 km/sec. The channel maps reveal a systematic trend kinematically, where the flux of [OIII] increases towards $v\sim 0$ km/sec in the regions outside the primary galaxy.

\begin{figure}
\includegraphics[width=3.25in]{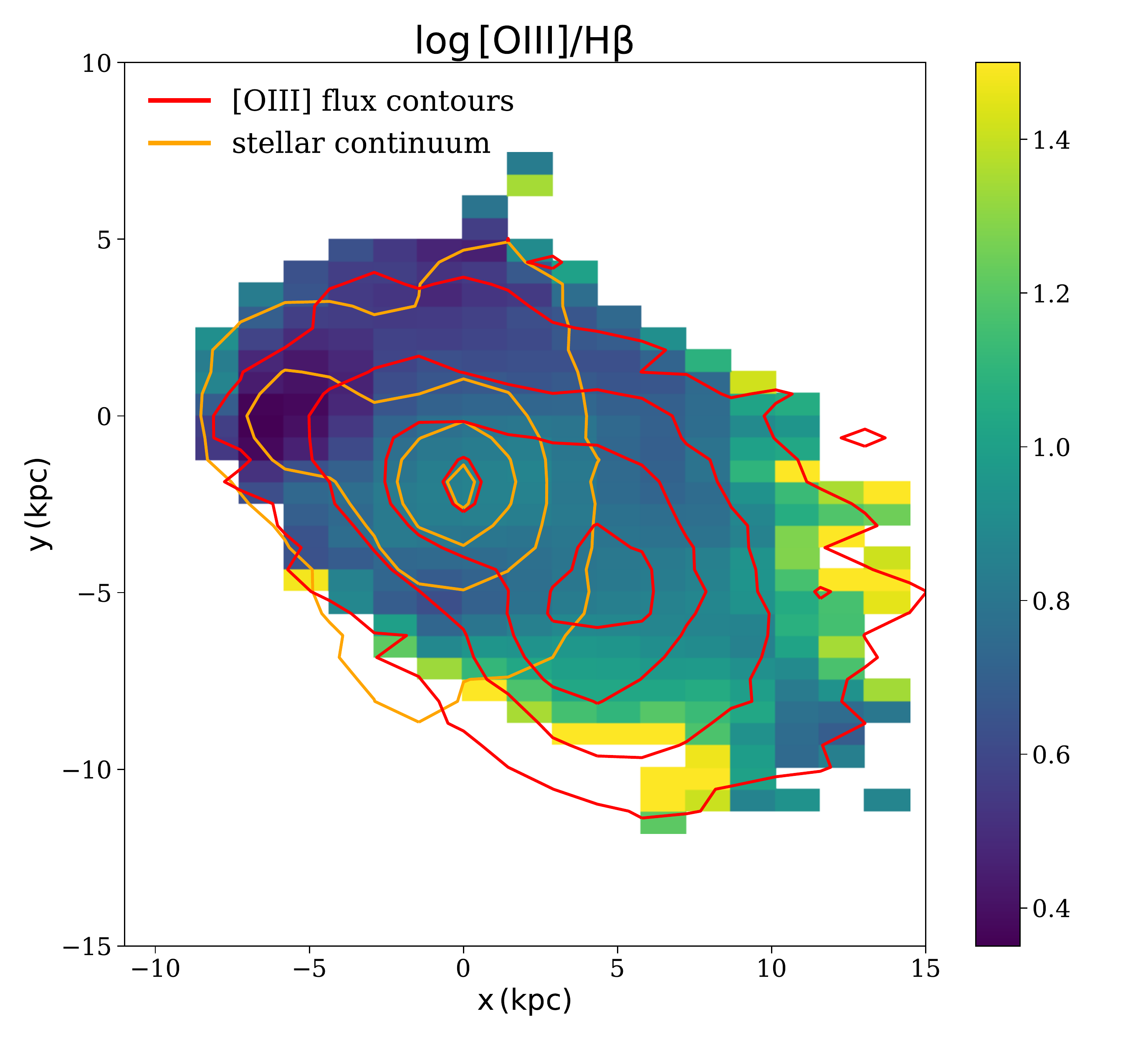}
\caption{$\mathrm{log\,[OIII]/H\beta}$ line ratio throughout the field. Orange contours represent the stellar continuum, where the largest contour represents the regions in which the continuum is not detected. The contours trace both the primary galaxy, at (0, 0) kpc, and the secondary galaxy, at (-5, 0) kpc. The red contours represent the [OIII] flux throughout the field. The secondary galaxy shows $\mathrm{log\,[OIII]/H\beta} \sim 0.4$, consistent with either a LINER or star forming system. The gas in the primary galaxy and outside it shows $\mathrm{log\,[OIII]/H\beta} \geqslant 0.8$, associated with Seyfert type emission.}\label{f:oiii_hbeta_ratio_image}
\end{figure}

We study the $\mathrm{log\,[OIII]/H\beta}$ ratio throughout the field. For each spaxel where both H$\beta$ and [OIII] are detected, we fit a single Gaussian to each emission line and measure its flux. We show this ratio in figure \ref{f:oiii_hbeta_ratio_image} using a continuous color-coding, indicated by the colorbar. The orange contours represent the stellar continuum emission, where the largest contour is the region beyond which the continuum emission is consistent with zero. The red contours represent the [OIII] emission line flux, with the largest contour marking the region beyond which [OIII] is no longer detected. We note that in the central spaxels around the primary galaxy the [OIII] line shows both broad and narrow components, but individual spaxels lack the necessary SNR to show both components in the H$\beta$ line. Since we fit single Gaussians to the lines, the spaxels in central regions show $\mathrm{[OIII]/H\beta}$ which is a combination of the two components found in the global spectrum (figure \ref{f:1NLR_BPT_diag}). Beyond about 2 kpc, we only find narrow components in both [OIII] and H$\beta$. This is due to the seeing during the KCWI observation and the lack of sufficient SNR in individual spaxels. Broad components are seen in both [OIII] and H$\beta$ to much larger distances, once the spaxels are summed. One can see in figure \ref{f:oiii_hbeta_ratio_image} that the extended gas (outside the primary galaxy) shows $\mathrm{log\,[OIII]/H\beta} \geq 0.8$, which is consistent with a Seyfert-like ionization source.

\begin{figure}
\includegraphics[width=3.25in]{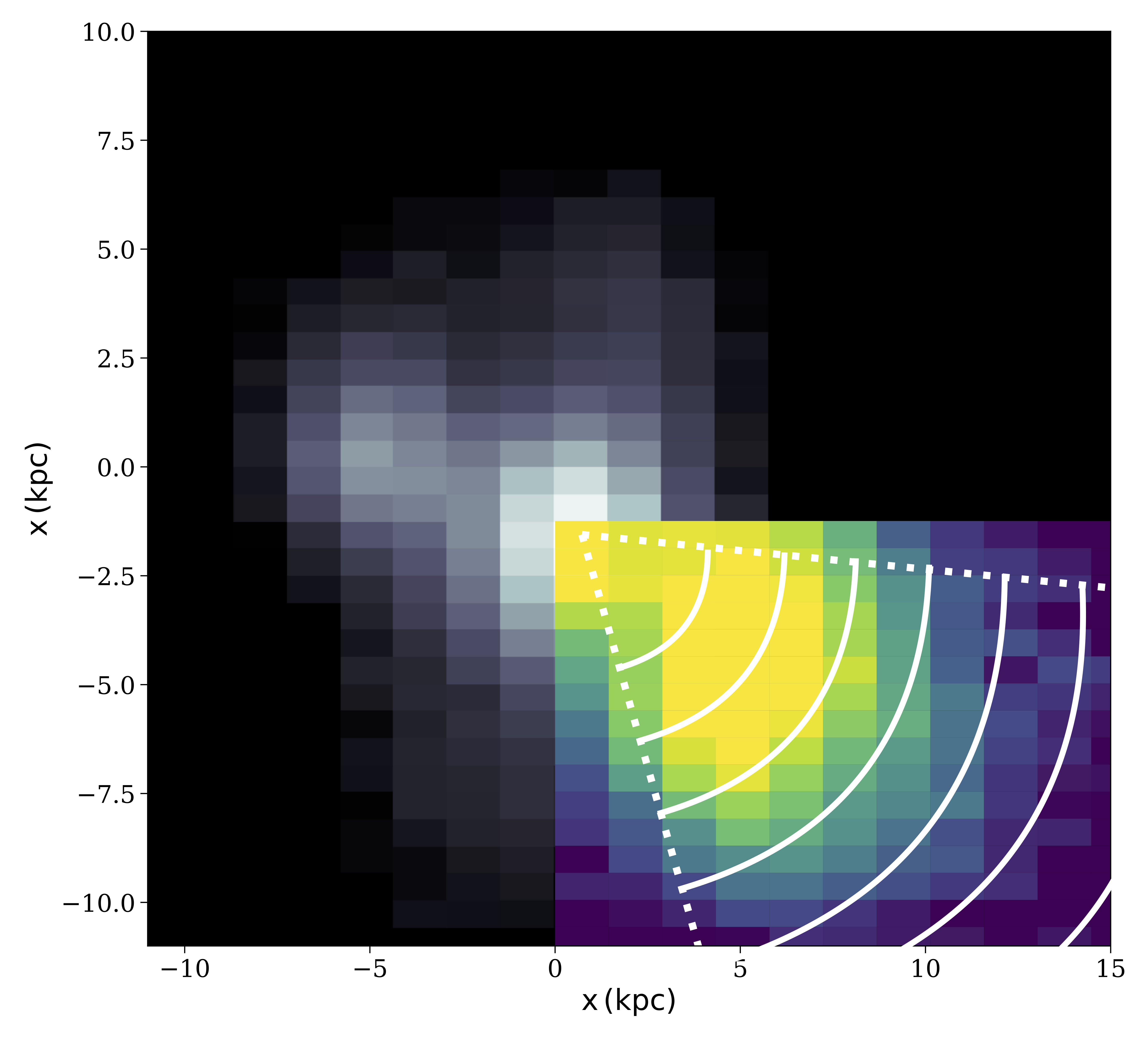}
\caption{Representation of the spatial bins used to study the gas cone outside the primary galaxy. The background color corresponds to the stellar continuum emission (in logarithmic scale) and traces the primary and secondary galaxies. The bottom right insert shows the [OIII] emission outside the galaxy with logarithmic color coding. The dashed lines mark the boundaries of the "gas cone" discussed in the text and the solid white lines the spherical shells used to obtain integrated line emission (the actual boundaries are slightly different and determined by individual pixels).}\label{f:layers_for_stacks}
\end{figure}

\begin{figure*}
\includegraphics[width=0.9\textwidth]{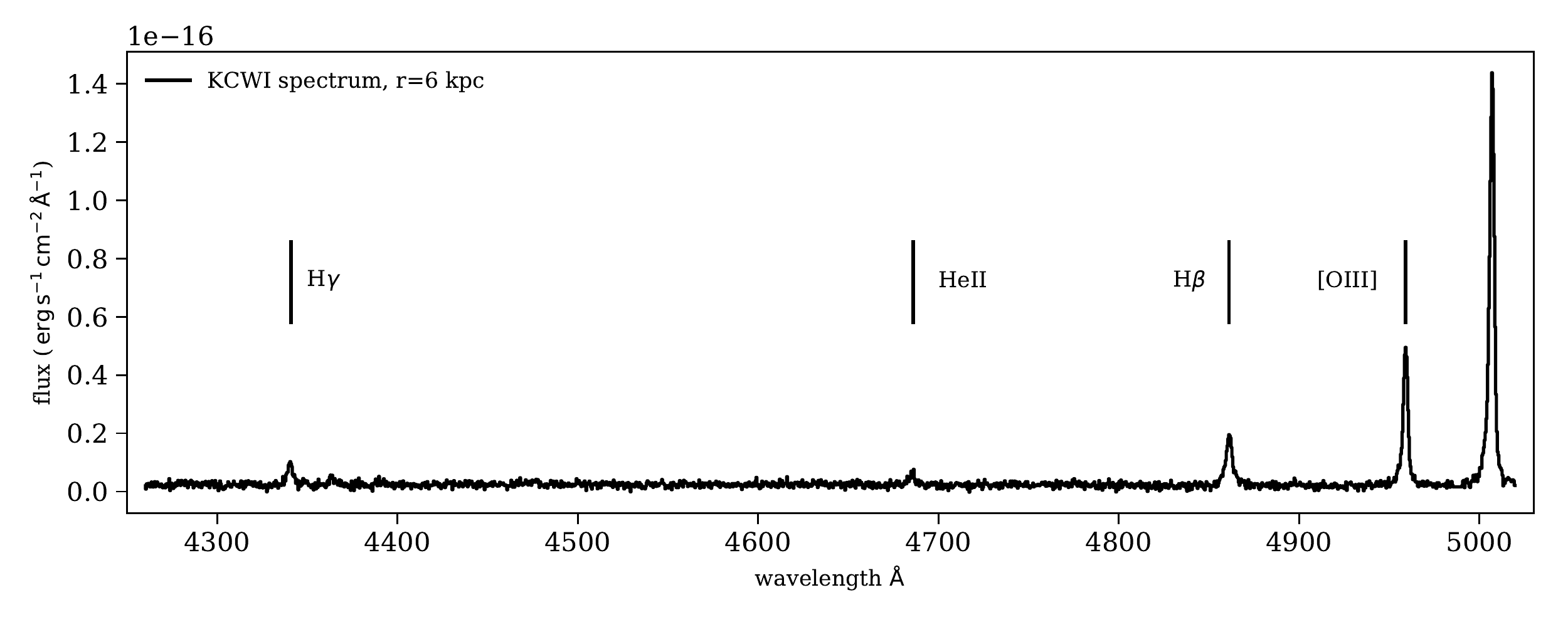}
\caption{An example of a combined KCWI spectrum of the region between 5 and 7 kpc along the gas cone. The spectrum is approximately associated with the red-colored spaxels in figure \ref{f:layers_for_stacks}. The combined spectra of other shells are of similar quality.}\label{f:kcwi_shell_spec_6kpc}
\end{figure*}

To study the extended gas properties as a function of distance from the center of the primary galaxy, and in order to increase the SNR, we sum spaxels with the same physical distance from the central galaxy. We define a cone, in which we find most of the extended gas emission, and mark its limits with dashed white lines in figure \ref{f:layers_for_stacks}. The opening angle of the cone is 70$^{\circ}$. We consider only spaxels that are between these two limits. We divide the region within the cone into seven shells, indicated by different white lines in figure \ref{f:layers_for_stacks}. We sum the spaxels within a given shell and obtain spectra as a function of distance from the center, where we take the median distance within each shell as the distance of the shell from the center. To appreciate the overall quality of the location-specific spectra, we show in figure \ref{f:kcwi_shell_spec_6kpc} the entire KCWI spectrum of the region between 5 and 7 kpc along the gas cone.

\begin{figure*}
\includegraphics[width=0.9\textwidth]{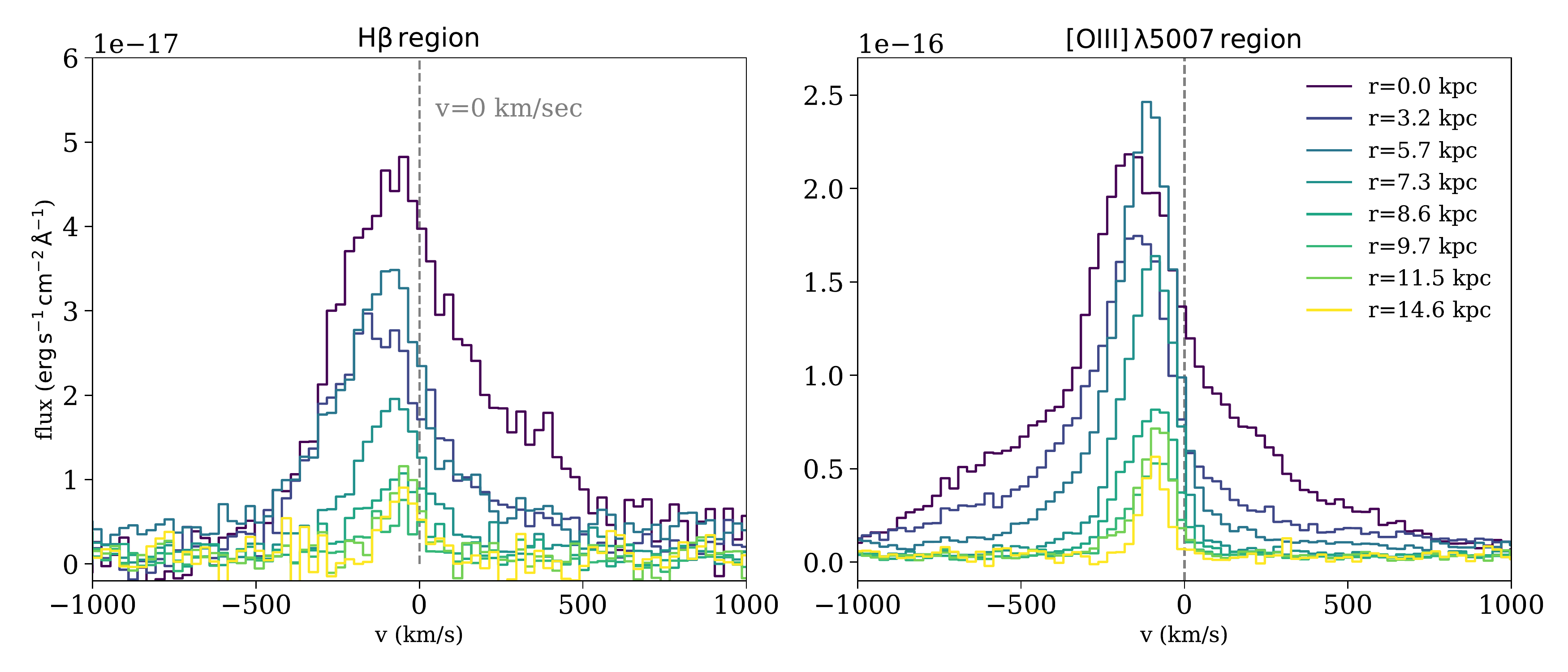}
\caption{$\mathrm{H\beta}$ (left panel) and [OIII] (right panel) flux density as a function of distance from the primary galaxy. These spectra were obtained by summing pixels within a certain distance from the primary galaxy. The x-axis shows the velocity, compared to systematic velocity of the stars in the primary galaxy. For distances of 0-7.3 kpc, blue and red wings are detected in the [OIII] emission profile. The blue wing is detected in [OIII] up to a distance of 11.5 kpc. }\label{f:oiii_and_hbeta_per_stack}
\end{figure*}

In figure \ref{f:oiii_and_hbeta_per_stack}, we show the [OIII] and H$\beta$ profiles of the seven shells and of the primary galaxy. The summed spectra show show both red and blue wings in their profiles for distances of 0-7.3 kpc, and an [OIII] asymmetry detected up to 11.5 kpc. As noted previously, we find large velocities in the [OIII] line, from approximately -1000 km/sec to +700 km/sec in the central regions of the galaxy. The [OIII] and H$\beta$ show asymmetric line profiles which cannot be modeled with a single Gaussian. The SNR in all seven shells is high enough to integrate over the profiles to obtain the line flux. We also detect H$\gamma$ and HeII~$\lambda$ 4686\AA\, emission within the cone, with lower SNR. The H$\gamma$ line profile can be described by a single Gaussian, therefore we fit both H$\beta$ and H$\gamma$ with single Gaussians, tying their widths and central wavelengths to have the same systematic radial velocity and velocity dispersion. The H$\beta$ is included in this fit to gain better constraints in modeling the H$\gamma$ line, and we find similar H$\beta$ fluxes when simply integrating over the flux and when using a single Gaussian fit. We show the best-fitting Gaussian profiles in figure \ref{f:hbeta_hgamma_fit} in appendix \ref{a:app}. The HeII line is even weaker than the H$\gamma$. In order to increase the SNR, we sum every pair of spectra, resulting in four (instead of eight) bins. We then fit simultaneously HeII and H$\beta$, each modeled with a single Gaussian, where we tie their central wavelengths and widths. The best-fitting profiles are shown in figure \ref{f:hbeta_helium_fit} in appendix \ref{a:app}.

\begin{figure*}
\includegraphics[width=0.9\textwidth]{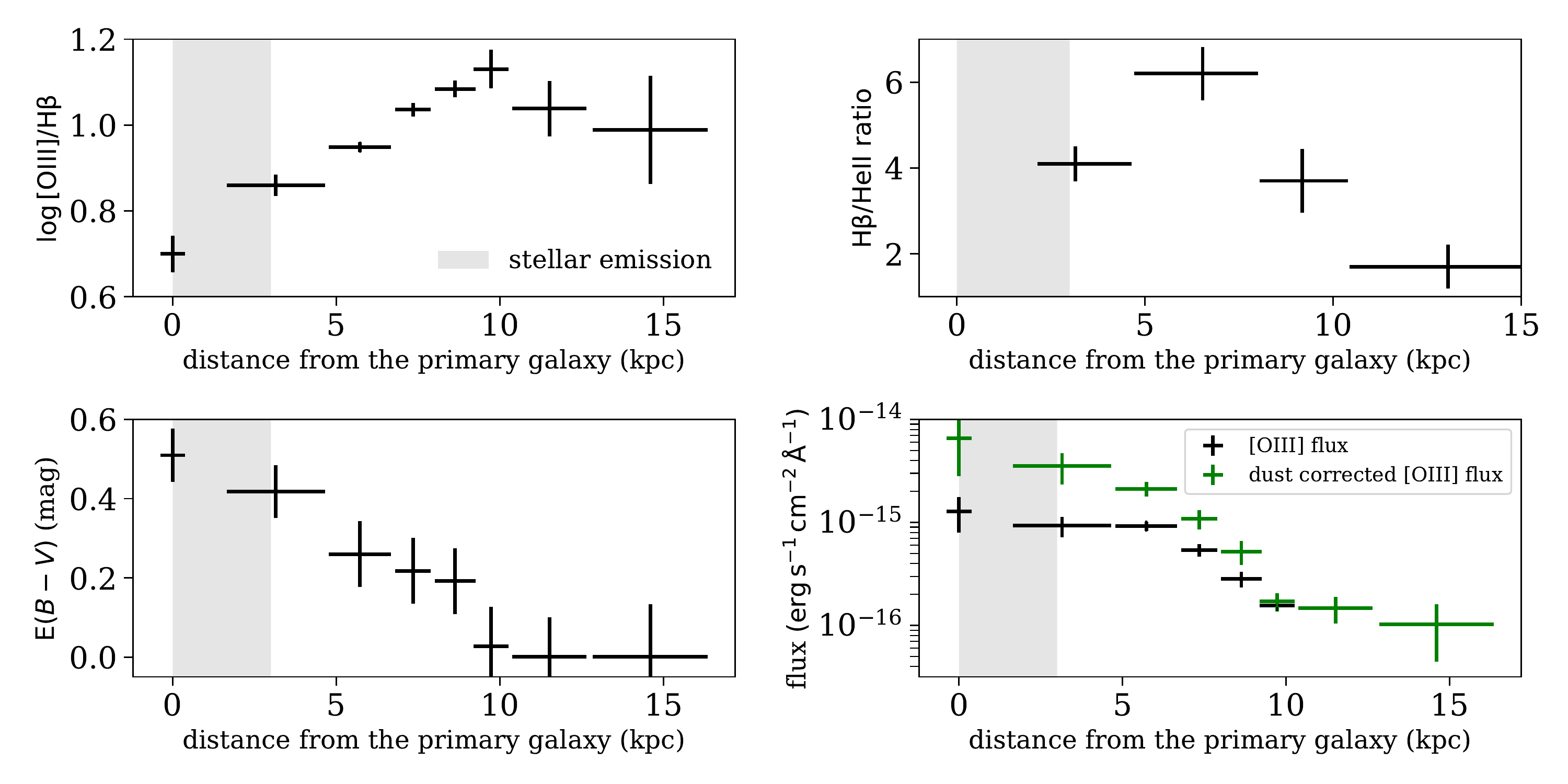}
\caption{Emission line properties as a function of distance from the central galaxy. The upper left panel shows the $\mathrm{[OIII]/H\beta}$ ratio from the primary galaxy and through the gas cone. The upper right panels shows the $\mathrm{H\beta/HeII}$ ratio throughout the gas cone, where we use fewer bins to increase the SNR of the HeII line. The lower left panel shows the $\mathrm{E}(B-V)$ extracted from the $\mathrm{H\beta/H\gamma}$ ratio under the assumption of a dusty screen, and the lower right panel shows measured and dust corrected $\mathrm{[OIII]}$ flux in the primary galaxy and throughout the gas cone, assuming a dusty screen.}\label{f:emission_line_diagnostics_blob}
\end{figure*}

Using the best-fitting profiles for the emission lines, we measure their flux as a function of distance from the central galaxy. The upper left panel of figure \ref{f:emission_line_diagnostics_blob} shows $\mathrm{[OIII]/H\beta}$ throughout the gas cone. The upper right panel shows the H$\beta$/HeII ratio. The ratio increases as a function of distance for all parts of the cone beyond 6 kpc. We can also estimate the dust reddening as a function of distance from the central galaxy using the H$\beta$ and H$\gamma$ emission lines. Assuming case-B recombination, gas temperature of $10^4$ K, a dusty screen, and the \citet{cardelli89} extinction law, the colour excess is given by:
\begin{equation}\label{eq:2}
	{\mathrm{E}(B-V) = \mathrm{4.53 log\, \Bigg[ \frac{(H\beta/H\gamma)_{obs}}{2.132} \Bigg] }}
\end{equation}
where $\mathrm{(H\beta/H\gamma)_{obs}}$ is the observed line ratio. We show $\mathrm{E}(B-V)$ as a function of distance from the center in the lower left panel of figure \ref{f:emission_line_diagnostics_blob}. We note that this estimate is highly uncertain due to the weak H$\gamma$ emission, and is only valid for a dusty screen geometry. We elaborate on this point in section \ref{s:photoionization_model}, where we present a full photoionization model for this system. In the lower right panel of figure \ref{f:emission_line_diagnostics_blob} we show the [OIII] flux along the gas cone (black), and the dust-corrected flux (assuming a dusty screen) in green.

We also examine the gas kinematics inside and outside the primary galaxy. First, we estimate the escape velocity from the galaxy using the light profile of the galaxy from the SDSS images. The exact process is described in \citet{baron17b}, and we summarize it briefly here. We assume that the mass profile is similar to the light profile, which is justified by the fact that the spectrum is dominated by a single-age population of stars. We then assume that the gas mass is at most half of the stellar mass. This is an upper limit since the gas mass observed in E+A galaxies is much lower (e.g. \citealt{rowlands15}). It is also supported by the evidence provided in the next sections. We then measure the line-of-sight FWHM
of the escape velocity (see \citealt{agnello14}), which is 450 km/sec for our system.

\begin{figure}
\includegraphics[width=3.25in]{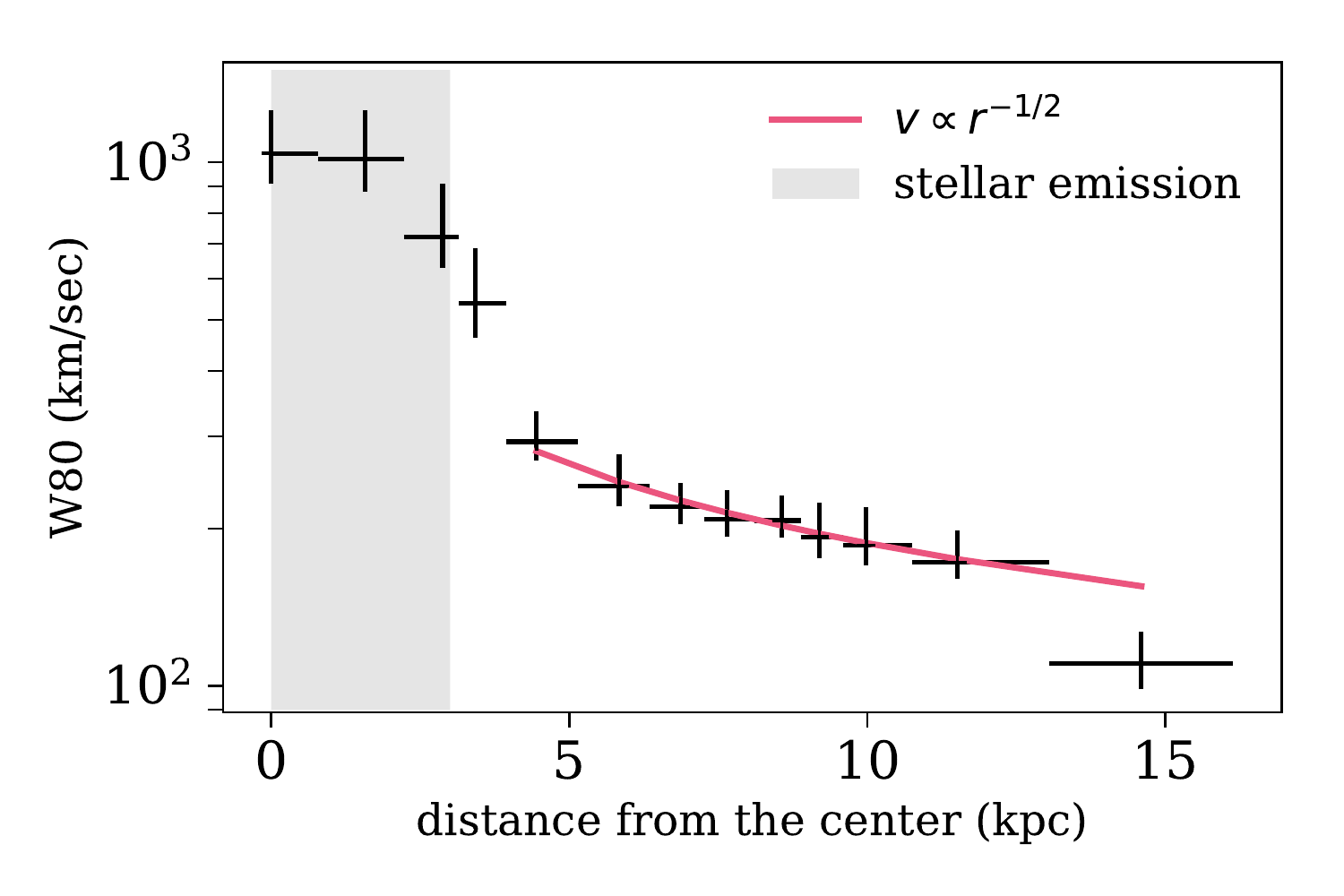}
\caption{$W_{80}\mathrm{[OIII]}$ as a function of distance from the center of the primary galaxy (black). The uncertainties on the x-axis represent the different shells we used, and the uncertainties on the y-axis represent the 85th and 75 percentiles of the flux. The gas at the outskirts of the galaxy follows a relation $v \propto r^{-1/2}$, which we mark with pink. We mark with grey the region where we detect stellar continuum emission from the primary galaxy. }\label{f:W80_versus_r}
\end{figure}

In figure \ref{f:oiii_and_hbeta_per_stack} we show that the [OIII] and H$\beta$ emission lines exhibit both narrow and broad components. The FWHM of the broad component exceeds the escape velocity from the galaxy in all the shells, thus we suggest that the ionized gas traces a large-scale galactic outflow, driven by the central BH. To study the radial dependence of the outflow in more detail, we use the highest SNR line of [OIII]. Since the line is strong enough, we divide the spaxels within the cone into 13, rather than 8, shells. For each shell, we measure $W_{80}$, which is defined as the width of the [OIII] that contains 80 percent of its integrated flux: $W_{80} = v_{90} - v_{10}$, where $v_{10}$ and $v_{90}$ are the velocities that correspond to the 10th and 90th percentiles of the integrated [OIII] flux. We choose to work with this definition in order to be consistent with previous studies of AGN-driven winds (e.g., \citealt{harrison14}) and to make sure that our measurements trace the broad rather than the narrow lines. For a continuous wind with spherical symmetry and a power-law radial dependence $v(r) \propto r^{-\alpha}$, $W_{80}$ will show the same radial dependence $W_{80} \propto r^{-\alpha}$, since the inclination of the system with respect to the observer will only change the normalisation of the measured velocity. Therefore, $W_{80}(r)$ traces the radial dependence of the wind in the system. In figure \ref{f:W80_versus_r} we show the measured $W_{80}$ for different shells (black). We find high gas velocities inside the galaxy (indicated by the grey area in the figure), and a strong drop in velocity as the gas goes out of the galaxy (at about 3-4 kpc). In most of the regions outside the primary galaxy, $W_{80}(r) \propto r^{-1/2}$, as indicated by the pink line. The latter is similar to the radial dependence of the escape velocity, assuming that all the mass is enclosed within 3--4 kpc and that the baryonic matter dominates to beyond 15 kpc.

\section{Modeling}\label{s:photoionization_model}

\subsection{BH properties}\label{s:AGN}

We show in section \ref{s:narrow_and_broad_comps} that the narrow lines within the primary galaxy are consistent with a LINER. Due to the strong resemblance of this system to the first system studied by us in \citet{baron17b}, in terms of gas properties, we argue that these emission lines are due to AGN photoionization rather than shock excitation or post AGB stars (see sections 4.1 and 4.2 there). The dust-corrected luminosities of [OIII] and [OI] were used to measure the bolometric luminosity of the AGN using the \citet{netzer09} expression:
\begin{equation}\label{eq:3}
	{\mathrm{log}\,L_{\mathrm{bol}} = 3.8 + 0.25\mathrm{log}L(\mathrm{[OIII]}) + 0.75\mathrm{log}L(\mathrm{[OI]})}
\end{equation}
This gives a bolometric luminosity of $L_{bol}=10^{44.5}\,\mathrm{erg\,s^{-1}}$ for this source. We note that the \citet{netzer09} relation is based on observations of many thousands of sources, on previous estimates of the luminosity (e.g. \citealt{heckman04}), and on detailed photoionization calculations. This relation describes the average source in the sample, and the scatter is at least 0.3 dex. We will return to this point in section \ref{s:photoionization_model}, where we construct a full photoionization model for the system.

The S$\mathrm{\acute{e}}$rsic index of the galaxy is 4.1 in the $r$-band and 4.9 in $g$-band, therefore the galaxy is bulge-dominated, and we can estimate the BH mass using the stellar velocity dispersion. We use the M-$\sigma$ relation from \citet{kormendy13} and find a BH mass of $\mathrm{log\,M/M_{\odot}=8.14\pm0.3}$, where we assume a nominal uncertainty of a factor 2 since we measure the velocity dispersion of the entire galaxy. For solar metallicity, $L_{\mathrm{Edd}}=1.5\times10^{38}\mathrm{(M_{BH}/M_{\odot})\,erg\,s^{-1}}$ thus, from the LINER component, $L/L_{\mathrm{Edd}}= 0.01$. This ratio is an order of magnitude smaller than what we found for the first E+A galaxy, and is more consistent with type-II LINERs than type-II Seyferts and type-I QSOs \citep{netzer09}. The corresponding accretion rate, assuming mass-to-radiation conversion efficiency $\eta_{acc}=0.1$, is $\dot{m}_{acc}=0.05\,\mathrm{M_{\odot}/yr}$.

\subsection{Photoionization modeling - general assumptions}\label{s:photo_1}

In section \ref{s:system_properties} we have shown that the primary galaxy is a post starburst E+A galaxy that harbours an active BH. The AGN presence is reflected through the emission line ratio [OIII]/H$\beta$ and the presence of a strong HeII emission, which suggests a hard ionizing continuum. We also found three ionized gas components: (1) the narrow line region, (2) a central unresolved broad kinematic component, with FWHM$\sim$1\,000 km/sec, and (3) gas that is located outside the primary galaxy and extends to a distance of 17 kpc, with a conic-like structure. We will refer to this component as the "gas cone". 

The presence of a luminous AGN, the line diagnostic diagrams and the lack of ongoing star formation, all suggest that the observed emission lines are due to photoionization of the gas by the central source. In this section we investigate the possible range of ionized gas and dust properties and present our best self-consistent photoionization models for the three components. This allows us to deduce the gas density, filling factor, covering factor, and metallicity. These properties are used later to compare our system with other type-II AGN and other E+A galaxies containing active BHs. It will also provide the necessary information to deduce the mass and mass outflow rate in all observed components.

All models presented below are designed to fit the following set of conditions:
\begin{enumerate}
\item The calculated H$\beta$, H$\gamma$, [OIII] and HeII line luminosities must agree with those observed in all the components at their various locations. They must also be consistent with the properties of the central source of ionization. 
\item The observed line ratios must be consistent with the location-dependent ionization parameter and the dust distribution in the source. 
\item The derived mass and mass outflow rates must be consistent with the gas density and filling factor derived in (i) and (ii) above. 
\item The derived dust extinction must be consistent with all the above as well as with the assumed metallicity of the gas. 
\end{enumerate}

The photoionization models presented below depend on various assumptions which contribute to the uncertainty of the derived gas properties, and the mass and mass outflow rate estimates. The main uncertainties in our models are:
\begin{enumerate}
\item The exact continuum SED.
\item The metallicity and distribution of dust within the gas.
\item The outflowing gas distribution. As detailed below, our most successful model assumes a conic-shaped continuous wind. However, we cannot distinguish this geometry from conic-shaped motion of a large number of small optically thin clouds with a small filling factor. 
\end{enumerate}
Given this, the optimization problem is not convex, and we cannot define these uncertainties and propagate them properly to derive well defined uncertainties. Nevertheless, the modeling choices we make introduce an uncertainty of at least 20\% to all the derived properties we list below.

\subsection{Photoionization modeling - the gas cone}\label{s:photo_gas_cone}
The gas cone provides most stringent constraints on the properties of the central source. Here and in what follows we use the standard definition of the ionization parameter,
\begin{equation}\label{eq:4}
	{\mathrm{U(r)} = \frac{\mathrm{Q(Lyman)}}{4\pi r^{2} n_{\mathrm{H}} c},}
\end{equation}
where $\mathrm{Q(Lyman)}$ is the number of hydrogen ionizing photons per unit time, $n_{\mathrm{H}}$ is the hydrogen number density, and $c$ the speed of light. We experimented with various ionizing SEDs, all based on a combination of thin accretion disk SED and a power-law X-ray continuum. The strong HeII line observed in the source suggests a "hard" ionizing source with a large fraction of ionizing photons above 4 Rydberg. The model chosen is consistent with a thin disk around $\mathrm{10^8}\,M_{\odot}$ BH with $L/L\mathrm{edd}=0.3$. The mean energy of an ionizing photon is 38 eV. The energy slope of the X-ray source is -0.75 and $\alpha_{OX}=1.38$, typical of the luminosity of the central sources.

The first comparison of the photoionization calculations for the gas in the cone with observed line luminosities assumes 90 degree inclination to the line of sight. Later, in section \ref{s:dyn_gas_cone}, we examine this assumption by comparing predicted and observed line profiles for a range of inclinations. The best fit inclination is very close to the one assumed here and the deviations from the numbers presented in this section are small.

The line ratios presented in figure \ref{f:emission_line_diagnostics_blob} constrain $\mathrm{U(r)}$ along the cone. The $\mathrm{[OIII/H\beta]}$ ratio is sensitive to the ionization state of the gas, its temperature and metallicity. The ratio ranges from 5 to 16, which sets tight constrains on the possible ionization parameters of the model to a range of $\mathrm{log\,U}=-2$ to $\mathrm{log\,U}=-0.5$. Given the estimated bolometric luminosity of the source (section \ref{s:agn_properties}), the constraint sets the relation between the hydrogen number density, $n_{\mathrm{H}}$, and the distance from the central source.

The $\mathrm{H\beta/HeII}$ ratio (upper right panel in fig. \ref{f:emission_line_diagnostics_blob}) gives an important clue to the ionization structure within the gas cone. The two recombination lines depend in the same way on the gas density and temperature, and their ratio does not depend on metallicity. We find that from $r\sim6$ kpc the $\mathrm{H\beta/HeII}$ ratio decreases, thus the He$^{++}$ fraction is expected to increase outwards. This can only be achieved if the ionization parameter increases outwards, which in turn requieres that the hydrogen density decreases faster than $r^{-2}$.

The luminosities of the observed emission lines $\mathrm{[OIII]}$, HeII, and $\mathrm{H\beta}$ as a function of distance from the central source (see, e.g., bottom right panel in fig. \ref{f:emission_line_diagnostics_blob}) set additional constraints on the model. While for a constant temperature, the line luminosities increase quadratically with the density, too large density and filling factor increase the gas opacity and result in a radiation bound configuration, where at large distances the gas becomes neutral and does not emit line radiation. Thus, the highly ionized strong emission line gas at large distances suggest a density bound system, where only part of the ionizing radiation is absorbed by the gas (i.e., the cone must be optically thin). 

Following these considerations, we model the gas cone as a single gas cloud that starts at a distance of 1 kpc from the central source and extends to 15 kpc, with density and filling factor that are modeled as simple power-laws with the distance. It is unclear where exactly is the transition beween gas that resides inside the galaxy and gas that is outside. Since the stellar continuum is no longer detected at about 3 kpc, we will compare the model to observations from 3 kpc to 15 kpc. We take the covering factor to be 0.1, since the gas cone shows an opening angle of about $70^{\circ}$ (projected on the sky). 

As explained in section \ref{s:AGN}, the bolometric luminosity of the AGN, based on the [OIII], [OI] and H$\beta$ lines in the NLR, is estimated to be $L\mathrm{_{bol}}=10^{44.4}\,\mathrm{erg/sec}$. These relations represent the mean of a very large number of type-II AGN and the uncertainty on this number, as judged from the scatter in the relationship, is at least 0.3 dex (see \citealt{netzer09}). Therefore, we consider a range of luminosities between $L\mathrm{_{bol}}=10^{44}\,\mathrm{erg/sec}$ and $L\mathrm{_{bol}}=10^{45}\,\mathrm{erg/sec}$ with the SED described earlier.

Next, we consider the gas density and filling factor. For the initial density at 1 kpc we explored the range $n_{\mathrm{H}}=10 \,\mathrm{cm^{-3}}$ to $n_{\mathrm{H}}= 500\,\mathrm{cm^{-3}}$, and for the radial dependence $n_{\mathrm{H}} \propto r^{-2}$ to $n_{\mathrm{H}} \propto r^{-3.5}$ (a steeper density law will not give sufficient flux at large distances). We consider a filling factor at the illuminated face of the cloud in the range $f=0.003$ to $f=0.1$, and a radial dependence in the range $f \propto \mathrm{const}$ to $f \propto r^{2}$. These choices are motivated by the constraints of line intensity and line extent and the various parameters were allowed to change until finding the best agreement with the observations. We note that for a specific choice of $n_{\mathrm{H}}$ and its radial dependance, there is little freedom in the choice of $f$ and its radial dependence. The two parameters are highly correlated due to the fact that the line luminosity depends on the product $fn_{\mathrm{H}} ^2$. Therefore, the range in radial dependence of the filling factor is set by the density power-law. One can expect a rising power-law in the filling factor in a case where gas in the galaxy is swept by the propagating wind. 

Finally, we experimented with different amounts of dust absorption and extinction by changing the gas metallicity from 0.3 to 3 times solar. Lacking additional constraints, we assume ISM-type grains with the relevant depletion from the gas phase. 

We run a grid of models with version 17.00 of \cloudy\ \citep{ferland17}. The parameters of the models vary within the ranges specified above, and the comparison with the observations was done in 7 shells that were defined by the observation described in section \ref{s:gas_properties} (fig. \ref{f:layers_for_stacks}). The comparison includes: [OIII], HeII, $\mathrm{H\beta}$, and $\mathrm{H\gamma}$. 

In section \ref{s:gas_properties} we estimated the dust reddening along the gas cone using the $\mathrm{H\beta/H\gamma}$ ratio, assuming a dusty screen. However, the dust used in the photoionization model is mixed with the gas. This will result in different attenuation of the emitted radiation partly because the emitting gas is spread throughout the entire volume and partly because such dust will only absorb radiation and will not scatter it out of the line of sight. Therefore, the two geometries give different $\mathrm{H\beta/H\gamma}$ ratios for a given amount of dust. To untangle this complex situation, we first extract the \emph{emergent} line luminosities of the model. These luminosities are the result of dust absorption within the gas cone. We use the \emph{emergent} $\mathrm{H\beta/H\gamma}$ line ratio and compare it to the observed one. We then added a dusty screen between the gas cone and the observer if the \emph{emergent} line ratio require such a component. The additional screen added in this way must be associated with neutral gas that is not exposed to the central source of radiation. We compute the value of $\mathrm{E}(B-V)$ by matching the model results with the observations. We then compare these luminosities to the observed luminosities in different shells. 

\begin{figure*}
\includegraphics[width=0.9\textwidth]{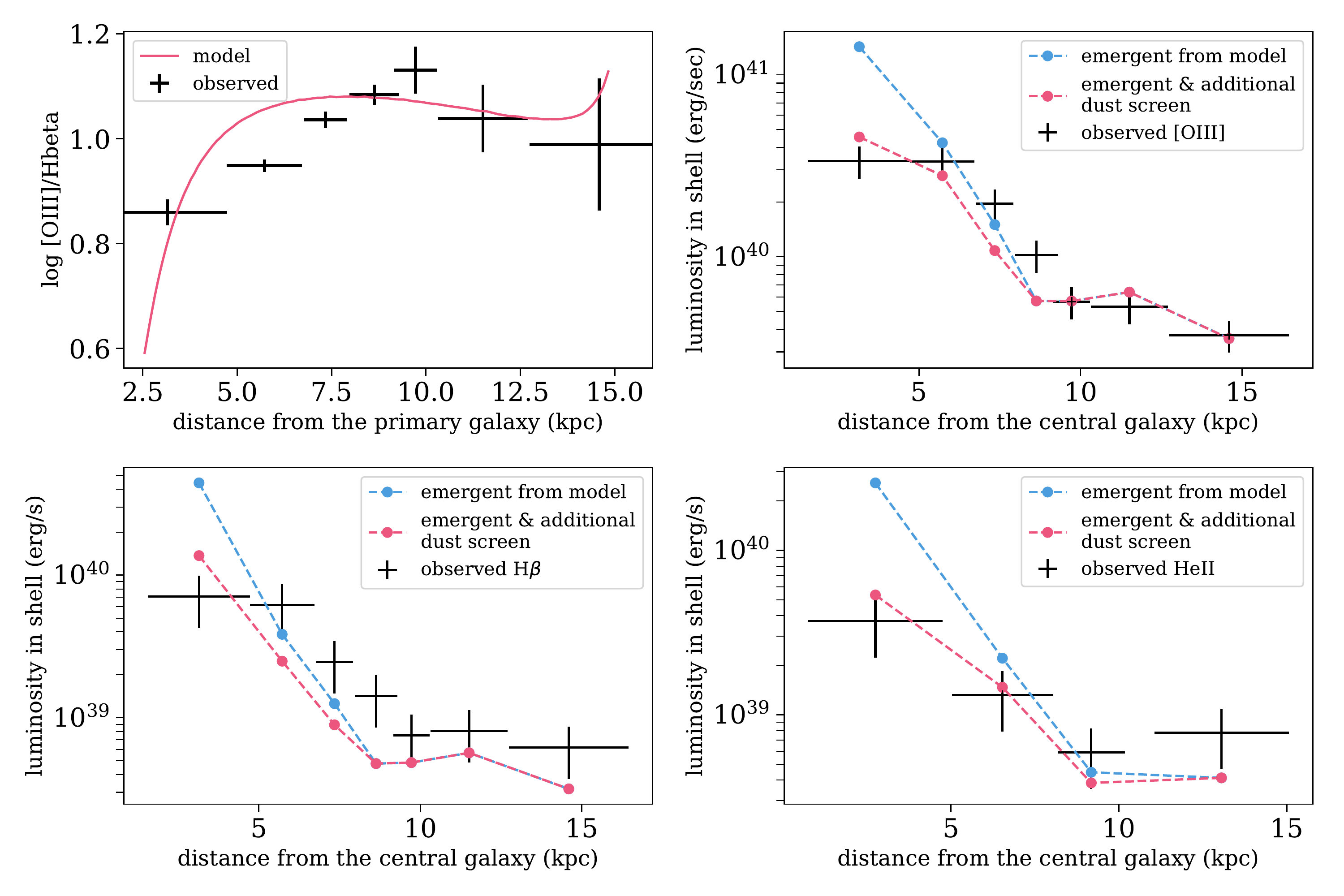}
\caption{Comparison of various observed properties of the gas cone to the best photoionization model. The model is composed of a central source with a bolometric luminosity of $\mathrm{L_{bol}=10^{45}}$ erg/sec, and with the SED described in the text. The gas has a spherical symmetry and extends from 1 to 15 kpc, with a covering factor of 0.1. The density is $n_{\mathrm{H}} = \mathrm{160\,(\frac{r}{1 kpc})^{-3}\, cm^{-3}}$ and the filling factor $f = \mathrm{0.005\,(\frac{r}{1 kpc})^{1.4}}$. The metallicity is half solar with ISM-type grains. The upper left panel compares the observed $\mathrm{log\,[OIII]/H\beta}$ ratio (black crosses) to that of the model (red line). The rest of the panels compare the observed line luminosities (no dust correction; black crosses) to the \emph{emergent} line luminosities before (blue) and after (red) applying additional reddening due to a dusty screen with $\mathrm{E}(B-V)$ obtained from the observed $\mathrm{H\beta/H\gamma}$ ratios.}\label{f:best_photoionization_model}
\end{figure*}

Given all that, the parameters of the best-fitting model are a central source with a bolometric luminosity of $L\mathrm{_{bol}}=10^{45}\,\mathrm{erg/sec}$, and SED typical of a high accretion rate thin disk, as explained earlier. The density in the model is $n_{\mathrm{H}} = \mathrm{160\,(\frac{r}{1 kpc})^{-3}\, cm^{-3}}$, the filling factor is $f = \mathrm{0.005\,(\frac{r}{1 kpc})^{1.4}}$, and the metallicity is half solar with ISM-type grains. We show in figure \ref{f:best_photoionization_model} the observed $\mathrm{[OIII]/H\beta}$ ratio compared to the predicted ratio (upper left panel), and a comparison between the observed line luminosities to the luminosities predicted by the model. In all three lines, we show both the \emph{emergent} luminosity (i.e., after accounting for internal reddening within the cloud; blue) and the \emph{emergent} luminosity after additional extinction by a dusty screen. Clearly, our model reproduces all the observations quite well. The exception is H$\beta$ line luminosity, which is systematically too low in the model.  

The model chosen here is quite different from several other models that did not provide a satisfactory solution. Some of these failing models succeed in predicting the [OIII] luminosity and failed for the $\mathrm{H\beta}$ and HeII luminosities (by more than an order of magnitude). Others result in an almost neutral cloud, with very low [OIII] and HeII luminosities and with acceptable $\mathrm{H\beta}$ luminosities. We also note that the best model in which an additional dusty screen (which is not part of the gas cone) is not necessary has a bolometric luminosity of $L\mathrm{_{bol}}=10^{45.4}\,\mathrm{erg/sec}$ and a density of $n_{\mathrm{H}} = \mathrm{275\,(\frac{r}{1 kpc})^{-3}\, cm^{-3}}$. This model gives H$\beta$ line luminosity which is consistent with observations, but fails to reproduce the line luminosities observed in the NLR, which is described in the following section. 

\subsection{Photoionization modeling - the NLR}

Next we model the NLR gas using the luminosity and SED obtained from the best fit model of the gas cone. The narrow lines are consistent with a LINER, and the emission line ratios we observe (see figure \ref{f:1NLR_BPT_diag}) require an ionization parameter of $\mathrm{log\,U \approx -3.6}$ for the chosen SED. We choose a thin gas slab, with a density of $n_{\mathrm{H}} = 10^{4} \, \mathrm{cm^{-3}}$ at a distance of 1 kpc, which reproduces the observed emission line ratios to within 0.2 dex. Since the NLR is not fully resolved by the observations, this combination of $\mathrm{n_H}$ and $r$ is not unique. The covering factor that is necessary to match the observed and the predicted line luminosities is 0.042. For the (rather arbitrary) chosen column density of $\mathrm{10^{21.5}\, cm^{-2}}$, the amount of reddening matches the observations and there is no need for an additional dust screen. 

All the properties noted above are consistent with observations of NLR in type II AGN (see \citealt{netzer09}, \citealt{heckman14}), with other E+A galaxies showing AGN-driven winds \citep{baron17b}, and with luminous LINERs observed in the local universe \citep{povic16}. 

\subsection{Photoionization modeling - the high velocity central component}\label{s:photo_broad}

Finally, we model the broad kinematic component in the central part of the galaxy, using the luminosity and SED of the best fit model of the gas cone. The broad lines are consistent with a Seyfert type spectrum (see figure \ref{f:1NLR_BPT_diag}), with an ionization parameter of about $\mathrm{log\,U=-2.4}$. We model the broad kinematic component as a thin shell at a distance of 0.85 kpc (see below), with a density of $n_{\mathrm{H}} = 10^{3}\, \mathrm{cm^{-3}}$. This choice reproduces the emission line ratios to within 0.2 dex, and for a covering factor of 0.3 it also reproduces the observed emission line luminosities. A covering factor of 0.3 is rather large for an NLR-type gas (see \citealt{netzer09}), but is consistent with an outflow with a large opening angle. It is also consistent with the covering factor found for the outflowing component in \citet{baron17b}, which is 0.5. 

As explained in section \ref{s:gas_properties}, the broad kinematic component is either unresolved or marginally-resolved, with a diameter that is not larger than $\sim$1 kpc. While we do not have a strong constraint on the location of this gas, the required ionization parameter sets a tight constraint of $\sim 6.2\times 10^{45}\,\mathrm{cm^{-1}}$ on the product $n_{\mathrm{H}}r^{2}$. There is a range of possible distances and gas densities that satisfy this constraint, and each of these will produce a model that is consistent with the observations. The range, 0.1-1 kpc, represents a real uncertainty on the mass and mass outflow rate of this component. We will use these numbers when estimating these properties in section \ref{s:mass_broad} below.

\subsection{Dynamical modeling - the gas cone}\label{s:dyn_gas_cone}

We now model the dynamics of the gas in the gas cone. Given the PSF, we focus on lines of sight with a projected distance larger than 5 kpc. In section \ref{s:gas_properties} (figure \ref{f:W80_versus_r}) we found that $W_{80}(r) \propto r^{-1/2}$, which means that the velocity of the wind is given by $v(r) = v_{0}\cdot \Big(\frac{r}{1\,\mathrm{kpc}}\Big)^{-1/2}$. We assume the same geometry as in the photoionization model - a cone with an opening angle of $70^{\circ}$, ranging from 0 to 17 kpc. We assume that the velocity, emissivity and density depend only on the distance from the center and use the line emissivity and gas density from the best-fitting photoionization model presented in section \ref{s:photo_gas_cone}. For a given line of sight, the emission line profile will depend on the inclination of the cone with respect to the line-of-sight. For a dustless gas and an inclination of $90^{\circ}$, the emission line profile is a double-horn symmetric profile around $v=0$ km/sec, while for an inclination of $0^{\circ}$ (looking into the cone) the entire line profile is blueshifted with respect to the systematic velocity, assuming the turbulent velocity is smaller than the radial velocity. For a dusty gas, the dust absorbs the emission lines originating in the farthest side of the cone more than in the closer side (with respect to the observer), as we discuss below.

Our goal is to fit the observed emission line profiles as a function of projected distance from the galaxy, and extract the best-fitting $v_{0}$ and inclination. We use the highest SNR line, [OIII]$\lambda$5007\AA, and examine five bins of projected distances, from 5 to 15 kpc. Generally, there are two angles which affect the resulting image, one of which is directly observed (see figure \ref{f:oiii_velocity_map}), and it is roughly 45 degrees with respect to the west-east axis. We therefore rotate the coordinate system so that the inclination with respect to this angle will be zero. We therefore have a single inclination angle to fit. 

We start from five regions that are observed as five partial thick shells. Given an inclination with respect to the line-of-sight, we project them back onto the real cone. This results in five regions inside the cone which are not spherical shells. Since the model provides the line emissivity at each location within the cone, all points inside the projected shells can be combined to give the predicted line emission, prior to dust extinction, in all five regions. The velocity of the gas within each projected shell is projected back onto the line-of-sight to obtain the emission line profile, which we compare with observations. Each projected shell contains dusty-gas, thus we need to take into account the dust extinction when constructing the line profile. The dust is distributed within the cone with the same radial dependence as the gas density (which we take from the photoionization model). For large enough inclinations, the dust will absorb the redshifted part (gas which is moving away) more than the blueshifted part (gas which is moving towards the observer). We use the gas density from the photoionization model and for each region within a given projected shell, we compute the gas column density from this region to the edge of the cone. We use this column density to determine the dust reddening of each such region. We apply the dust reddening to the intrinsic emissivity of each region before adding it to the combined line profile. Finally, we convolve the resulting emission line profile with a Gaussian with a standard deviation of $\sigma_{turb}$, representing the (unknown) turbulent velocity in the gas.

\begin{figure*}
\includegraphics[width=0.9\textwidth]{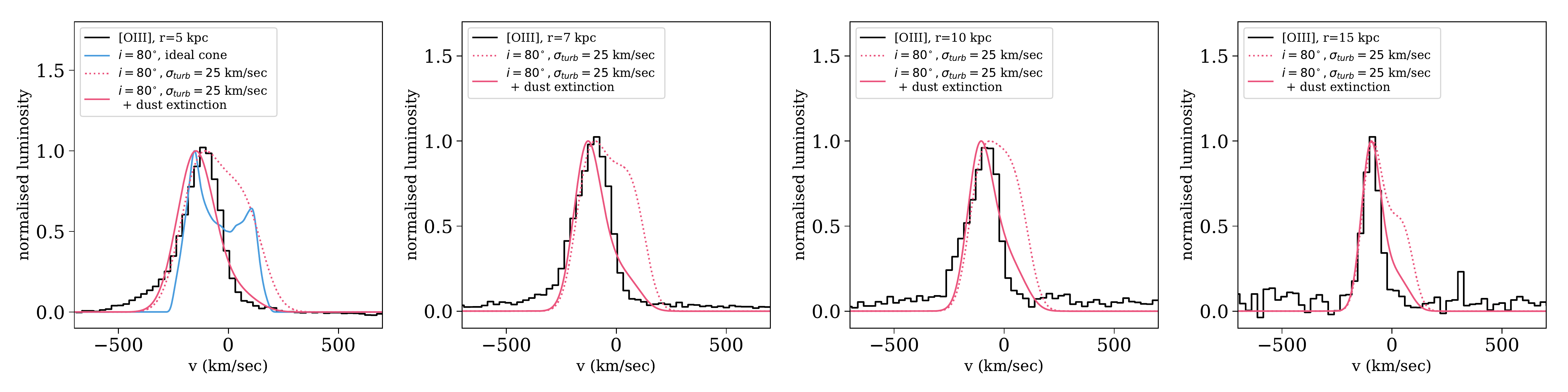}
\caption{Comparison of the observed [OIII] emission line profile (black) with a dynamic model of an expanding gas cone. The gas in the cone is expanding with a velocity of $v(r) = 800 \cdot \Big( \frac{r}{1\, \mathrm{kpc}} \Big) ^{-1/2}$ km/sec, and its emissivity and density are taken from the best-fitting photoionization model. The inclination of the cone with respect to the LOS is $i=80^{\circ}$. The left panel shows an ideal cone (without turbulence and dust extinction) with the distance-dependent emissivity, which we mark with blue. The other panels show the emission line profile before and after applying reddening along the line of sight, which we mark with dotted and solid pink lines, respectively. The velocity dispersion due to turbulent motion is $\sigma_{turb} = 25$ km/sec in all cases.}\label{f:cone_velocity_fit}
\end{figure*}

The result of this procedure is an emission line profile, which depends on the distance and on three free parameters: (1) $v_{0}$ - the wind velocity at 1 kpc, (2) $i$ - the inclination of the gas cone with respect to the observer, and (3) $\sigma_{turb}$ - the turbulent velocity dispersion in the gas. We show in figure \ref{f:cone_velocity_fit} such a model, with $v_{0}=800$ km/sec, $i=80^{\circ}$, and $\sigma_{turb}=25$ km/sec. We mark the observed [OIII] emission line with black, and the modeled emission line profile with pink, before and after applying the dust reddening along the line of sight (dotted and solid lines respectively). We find reasonable fits to the emission line profiles for $v_{0}$ in the range 700--900 km/sec, $i$ in the range 75-85$^{\circ}$, and turbulence velocity dispersion of 20--30 km/sec. As clearly seen in the diagram, dust attenuation of the red wing of the line in all shells is key to the good agreement between model and observations.

As noted earlier, the best-fitting photoionization model for the gas cone presented in section \ref{s:photo_gas_cone} is calculated under the assumption that the inclination of the cone with respect to the observer is $90^{\circ}$. However, we find that for inclinations in the range $70-90^{\circ}$, the projection of the best-fitting photoionization model changes the emission line luminosities and ratios by no more than 0.05 dex. Therefore, the best-fitting photoionization model is consistent with the best-fitting dynamic model of the wind.

\subsection{Mass and mass outflow rates}\label{s:masses}
The models presented above are detailed enough to deduce the mass and mass outflow rate in each of the four components: the gas cone, the neutral dusty screen, the NLR, and the central high velocity gas. Below we derive mass and mass outflow rate for each using the best fit photoionization models and the detailed velocity maps presented in figures \ref{f:oiii_velocity_map} and \ref{f:W80_versus_r}. The uncertainties on the mass and mass outflow rates are dominated by the uncertainties in our photoionization models, and are at least 20\% (see section \ref{s:photo_1}). The geometry of the emitting gas is modeled through the fit to the observed line profiles and luminosities and therefore does not add significantly to the overall uncertainty.

\subsubsection{The gas cone}\label{s:gas_cone_masses}

As explained in section \ref{s:photo_gas_cone}, we modeled the gas outside the galaxy as a cone with an opening angle of 70$^{\circ}$. The mass in the gas cone is given by:
\begin{equation}\label{eq:5}
	{M_{\mathrm{gas}} = 4 \pi \mu m_{\mathrm{H}} C_f \mathrm{\int_{3\,kpc}^{15\,kpc}  r^2 n_{H}(r) f(r) dr }}
\end{equation}
where $\mu$ is the mean molecular weight, $m_{\mathrm{H}}$ is the hydrogen mass, $C_f$ is the distance-independent covering factor (which is simply set by the opening angle of the cone), $n_{\mathrm{H}}(r)$ is the distance-dependent hydrogen density, and $f(r)$ the distance-dependent filling factor. The integrated mass to a given distance from the central galaxy is proportional to $r^{1.4}$. We choose the integration limits to be from 3 kpc (approximately where we no longer detect stellar continuum emission) to 15 kpc. The outer boundary of the detectable wind is probably somewhere between 15 and 17 kpc. Using $r=15$ kpc as the upper limit, and the density and filling factors obtained in section \ref{s:photo_gas_cone}, we get a gas mass of $M_{\mathrm{gas}} \approx 7\times 10^{8}\,\mathrm{M_{\odot}}$. For $r=17$ kpc we find $M_{\mathrm{gas}} \approx 10^{9}\,\mathrm{M_{\odot}}$. 

Our best-fitting photoionization model slightly underestimates the H$\beta$ emission line luminosity (see figure \ref{f:best_photoionization_model}). A model that does not underestimate the H$\beta$ luminosity (and hence will slightly overestimate the [OIII] luminosity) requires higher gas density and larger total mass. 

Assuming that the wind in the gas cone is continuous, the mass outflow rate is:
\begin{equation}\label{eq:6}
	{\dot M = 4 \pi \mu m_{\mathrm{H}} r^{2} C_f n_{\mathrm{H}}(r) f(r) v(r) = \mathrm{const}}
\end{equation}
where $v(r)$ is the wind velocity, assuming spherical symmetry, where the opening angle is not changing with the distance. Using the best-fitting power-laws from the photoionization model to extract the radial dependence of the wind velocity, we find $v(r) \propto r^{-0.4}$, very close to the relation we observe using the [OIII] emission line profile, $W_{80} \propto r^{-0.5}$ (see section \ref{s:gas_properties} and figure \ref{f:W80_versus_r}). Under these assumptions, the profiles inferred from the photoionization model are consistent with the observations. Alternatively, we can state that the photoionization model and the observed gas kinematics imply that the wind in the gas cone is a continuous flow. Below we use these numbers to put limits on the time that the BH is active (section \ref{s:bh_limit}), and to infer the fate of this outflowing gas (section \ref{s:fate}). Using the location-dependent gas density and filling factor from the best-fitting photoionization model, the mass outflow rate is $\dot M \approx 24 \, \mathrm{M_{\odot}/yr}$. We integrate from 3 to 17 kpc over the velocity of the gas and find that the kinetic power of the wind is roughly $8.5\times 10^{-3}L_{\mathrm{bol}}$.

\subsubsection{The neutral dusty screen}

We found in section \ref{s:photo_gas_cone} that in order to be consistent with the observations one must add a dusty screen between the gas cone and the observer. This is a key aspect of the kinematic modeling as well (see section \ref{s:dyn_gas_cone}). We require this screen to be neutral, as it does not contribute to the observed line emission. The additional dust reddening is needed in the first three shells, with $\mathrm{E}(B-V) = 0.3$ mag at 3.2 kpc, $\mathrm{E}(B-V) = 0.13$ mag at 5.7 kpc, and $\mathrm{E}(B-V) = 0.1$ mag at 7.3 kpc. For solar metallicity, the corresponding column densities are $\mathrm{N_{H}} = \mathrm{E}(B-V) \mathrm{\times 5.6 \cdot 10^{21}\, cm^{-2}}$. The minimal mass is obtained by adding three screens covering exactly the corresponding parts of the cone. This gives $M_{\mathrm{screen}} \sim 1.4 \times 10^{8}\, \mathrm{M_{\odot}}$. The maximum mass is roughly 2-3 times larger, depending on the other parts of the galaxy obscured by such a screen. Thus, the neutral screen mass is 2--5 times smaller than the mass in the cone albeit with a large uncertainty.

\subsubsection{The NLR}

For the NLR we assume a spherical thin shell geometry. The thickness of the shell is only limited by the requirement of a radiation bounded gas. For a column density of $N_{\mathrm{H, NLR}} = 10^{21.5}\, \mathrm{cm^{-2}}$, the mass is:

\begin{equation}\label{eq:9}
	{M_{\mathrm{NLR}} = 4\pi C_{f, \mathrm{NLR}} r_{\mathrm{NLR}}^2 N_{\mathrm{H, NLR}} \mu m_{\mathrm{H}}}
\end{equation}
where $C_{f, \mathrm{NLR}} = 0.042$ is the NLR covering factor, $r_{\mathrm{NLR}} = 1$ kpc, and $N_{\mathrm{H, NLR}} = 10^{21.5}\, \mathrm{cm^{-2}}$. The NLR gas mass is therefore $M_{\mathrm{NLR}} \approx 1.5 \times 10^7\, \mathrm{M_{\odot}}$.

\subsubsection{The central high velocity ionized component}\label{s:mass_broad}

The photoionization model presented in section \ref{s:photo_broad} is not unique because of the uncertainty on the location of the gas. The procedure we use to estimate the mass and the mass outflow rate of the ionized gas is similar to the one used by \citet{baron17b}, which is based on the known ionization parameter and hence the known value of $n_{\mathrm{H}} r^{2}$ ($6 \times 10^{45}\, \mathrm{cm^{-1}}$, see section \ref{s:photo_broad}). The ionized gas mass is directly related to the H$\alpha$ luminosity (see e.g. \citealt{soto12} and \citealt{baron17b}):
\begin{equation}\label{eq:10}
	{M_{\mathrm{out}} = \frac{\mu m_{\mathrm{H}}}{\gamma n_{\mathrm{H}}} L_{\mathrm{H}\alpha}}
\end{equation}
where $\gamma = 3.56 \times 10^{-25}\,\mathrm{erg\,cm^{3}\,s^{-1}}$ for case B recombination and $T_{e} \sim 10^{4}\,\mathrm{K}$ \citep{osterbrock06}, and $L_{\mathrm{H\alpha}} = 4.5\times 10^{42}$ erg/sec. This results in a mass estimate which depends only on the location of the gas:
\begin{equation}\label{eq:11}
	{M_{\mathrm{out}} \approx 2.5 \times 10^{7} \mathrm{\Big( \frac{r_{out}}{1\, kpc} \Big)^{2}\, M_{\odot} }}
\end{equation}
where $\mathrm{r_{out}}$ is in the range 100 pc to 1 kpc (see section \ref{s:photo_broad}). The upper limit on the mass, $2.5 \times 10^{7}\, \mathrm{M_{\odot}}$, is 50 times lower than the mass in the gas cone, with a kinetic power of roughly $1.5 \times 10^{-3}L_{\mathrm{bol}}$. A possible interpretation is that the AGN has already driven most of the primary gas outside of the galaxy. Obviously, the fast moving ionized gas can be radiation bounded, in which case the total mass of this component, and its kinetic energy, must include an additional contribution from neutral gas at the back side of the outflowing material.

The mass outflow rate for this component is given by $\dot M_{\mathrm{out}} = M_{\mathrm{out}}/ t_{\mathrm{out}}$, where $t_{\mathrm{out}} = r_{\mathrm{out}} / v_{\mathrm{out}}$. We take $v_{\mathrm{out}}$ as the velocity of the peak of the broad emission relative to the peak of the narrow, added to the width of the broad emission, $v_{\mathrm{out}} = v_{\mathrm{peak, broad}} + \sigma_{\mathrm{broad}} = 550\,\mathrm{km\,s^{-1}}$. The resulting outflow rate is:
\begin{equation}\label{eq:12}
	\mathrm{\dot{M}_{out}} \approx 16 \, \Big(\mathrm{\frac{r_{out}}{1\,kpc}}\Big)\, \mathrm{M_{\odot}/yr}.
\end{equation}
Taking $v_{\mathrm{out}} = W_{80}/1.3$ (see \citealt{harrison14}) instead, results in $v_{\mathrm{out}} = 830\,\mathrm{km\,s^{-1}}$, which increases the outflow rate by a factor of 1.5 to $\dot M \approx 24 \, \mathrm{M_{\odot}/yr}$, similar to the mass outflow rate in the gas cone. Given the limits on the winds location, we find that the mass outflow rate for this system is more than an order of magnitude lower than what we found in \citet{baron17b} for the central broad component in the E+A galaxy SDSS J132401.63+454620.6 (the maximum in that case was $120\, \mathrm{M_{\odot}/yr}$). Most of the difference between the two E+A galaxies is due to the fact that the broad H$\alpha$ luminosity in the present case is about an order of magnitude smaller compared to the case presented in \citet{baron17b}. Again, these calculations provide only lower limits since they do not take into account neutral gas at the back side of the flow.

\section{Discussion}\label{s:disc}

The observations and models presented in this paper link the central AGN with high velocity ionized gas flows in two regions: a central region with a distance of 1 kpc or less from the AGN, and a conical shaped region that extends up to about 17 kpc from the center of the primary galaxy. Our IFU measurements, and detailed dynamical and photoionization modeling of this gas, clearly indicate an AGN-driven flow. To the best of our knowledge, this is the very first time that continuous, wind-type flows, that are accurately timed (through the known stellar population), and with such well determined mass and mass outflow rates, are reported. It is also the clearest case, so far, showing that the amount of gas removed by the wind, is most of the gas remaining in the system after the quenching of SF. This seems to be a classical example of an AGN feedback where quenching is directly related to gas removal from the system.

Given the likely history of the system in question (merger and ULIRG) we must also consider two alternative scenarios related to the origin of the ionized gas outside the galaxy. We focus on the gas in the cone where the observational and modeling constraints are tighter and where the mass of the ionized gas is the largest. 

\subsection{Alternative origins for the gas in the cone}\label{s:origin_of_cone}

\subsubsection{Gas stripped from the companion galaxy}

One possible interpretation of the existence of a significant amount of gas outside the primary galaxy concerns the interaction itself. It is possible that the observed gas was stripped from the companion galaxy during its close pass near the primary galaxy, about 100--200 Myrs ago. According to this scenario, the stripped gas cooled down and expanded until the AGN turned on, and photoionized this gas. To explain the conic-shape emission line region, this gas is falling onto the primary galaxy. Following the details of the process requires numerical simulations and is beyond the scope of our paper.

An inflow with a large opening angle can, in principle, produce both blueshifted and redshifted emission lines, with respect to the systematic velocity of the primary galaxy. The cone-like structure is more difficult to explain but even this assumption runs into real difficulties. For an outflowing cone, the blueshifted part is produced by the closest region in the cone, and the redshift part by the farthest region in the cone. For an ideal cone, viewed at an inclination of 90$^{\circ}$, both of them produce a double-horn, symmetric around 0 km/sec, emission line profile. However, since the gas is dusty, the dust absorbs the emission lines originating on the far side more than on the near side. Therefore, for an inclination close to 90$^{\circ}$, for an outflowing gas cone, the dust will absorb the redshifted part of the emission line profile, while for an inflowing cone, the dust will absorb the blueshifted part of the profile. Figure \ref{f:cone_velocity_fit} clearly shows that the redshifted part is absorbed, while the blueshifted part is not. Therefore, dynamical considerations, based on the line profiles, rule out this explanation.

A cone with an inclination of close to 180$^{\circ}$, where the line of sight points straight to the in-falling gas, will produce a primarily-blueshifted emission. This scenario cannot account for the redshifted emission. Furthermore, this scenario requires large projection corrections, which result in an infall velocity which exceeds greatly the expected free fall velocity of the gas at a given distance from the primary galaxy.

\subsubsection{Starburst-driven outflow}

The observed gas outside the primary galaxy may also be due to supernovae-driven winds during the recent starburst. Such a scenario was suggested by \citet{heckman17} to explain possible differences of the CGM properties observed in 17 E+A post starburst galaxies, compared to a control sample of non-post starburst systems. During the starburst, supernovae-driven winds can accelerate gas to substantial velocities, which are large enough to remove the gas from the system (see \citealt{heckman17} and \citealt{heckman17b} for a detailed review). Such observed starburst-driven winds are multi-phased, where each phase differs in its temperature, density, and velocity. The warm-ionized phase, which is traced by optical emission lines, shows wind velocities of hundreds to a thousand km/s, without correcting for projection effects. Winds that are traced by interstellar absorption lines show roughly the same velocities (see \citealt{heckman17b} for a full discussion of properties and caveats). 

The best-fitting stellar population synthesis model for our system suggests a starburst that started 400 Myrs ago, and decreased exponentially to less than 10\% of its initial SFR values after 200 Myrs. Therefore, most of the core-collapse supernovae occurred roughly 300 Myrs ago. Assuming a typical outflow velocity of 500 km/sec, the gas would have traveled 150 kpc by now, far beyond the region where we detect the outflow. Alternatively, in order to detect supernovae-driven wind at a distance of 17 kpc from the center of the primary galaxy, the wind velocity must be around 50 km/sec, much smaller than observed startburst-driven winds, and more importantly, much smaller than the observed velocity of the wind (roughly 800 km/sec).

\subsection{AGN activity time}\label{s:bh_limit}

The gas properties inferred from the best-fitting photoionization model, combined with the observed gas kinematics, suggest that the observed wind forms a continuous flow. This in turn suggests that the AGN was active for the time it took the gas to reach 17 kpc. We use the inferred wind dynamics to put limits on the AGN activity time. We use the velocity law found earlier, $v(r) = v_{0} \times r_{kpc}^{-0.5}$, where $v_{0}$ is in the range 700--900 km/sec, integrate between 3 kpc and 17 kpc (the regions where the wind obeys the continuity equation), and obtained an AGN activity time of $\sim$60 Myrs. This value is a lower limit, since the AGN could have been active for longer time without launching a wind during the first part of the activity phase. In addition, the wind may exist at further distances beyond 17 kpc but is too ionized and/or with a too-low density to be detected with KCWI. An activity time of about 60 Myrs is in the general range considered in other estimates that range from about $10^6$ to $10^8$ years (see e.g. \citealt{haiman01} and \citealt{martini01}). It is much longer than the AGN flickering time-scale, of order $10^{5}$ yrs, suggested by several theoretical and observational studies (see e.g. \citealt{schawinski15} and \citealt{oppenheimer18}, and discussion by \citealt{sartori18}).

Given this, we infer that the AGN became active about 150 Myrs after the end of the starburst. We cannot examine the possibility of an earlier, perhaps similar AGN activity phase prior to the one considered here.

\subsection{Infrared emission, UV absorption, and the fate of the system}\label{s:fate}

Our photoionization model of the dusty gas in the gas cone suggests a massive gas component outside the primary galaxy. As long as the AGN remains active, this component should be observed through mid-infrared dust emission. According to our best-fitting photoionization model, the dust temperature ranges from 40--70 K at 1 kpc, to 10--20 K at 15 kpc, depending on the specific grain composition. The spectrum of this dust shows prominent mid-infrared emission, with a broad peak around $50 \mu m$. This peak emission is between the typical torus emission which peaks at 5--30 $\mu m$ and SF heated dust with a peak around 70--100 $\mu m$. Contrary to the dust in the torus, the infrared emission from the gas cone is expected to be spatially resolved.

The optical emission lines observed by the KCWI trace a specific gas phase of the wind, with a temperature in the range $10^{4}$--$10^{4.5}$ K. A hotter gas component, with a temperature of $10^{5}$--$10^{5.5}$ K can be traced by UV absorption lines, using background objects as continuum sources. Indeed, \citet{tripp11} used the COS instrument on the HST and found "hot-warm" plasma at $10^{5.5}$ K, traced by various UV absorption lines, at a projected distance of 68 kpc from a post starburst E+A galaxy. According to their modelling, this component contains 10--150 times more mass than the colder gas in a post starburst galaxy wind. Since the ionization parameter of the gas is increasing with the distance from the galaxy, and the gas temperature is expected to rise outwards, the gas may produce observable UV absorption lines. It may also become thermally unstable and reach much higher temperatures. 

Finally, we can extrapolate the gas density, inferred from the best-fitting photoionization model, and the observed gas dynamics to give a rough estimate for the fate of the system. It would take the gas about 1 Gyr to reach a distance of 150 kpc, at this stage the gas density would be $5 \times 10^{-5}\, \mathrm{cm^{-3}}$ and its velocity will be of order 50 km/s. Obviously, existing CGM can stall the wind well before it reaches 150 kpc. Regardless of the exact scenario, such a wind will contribute its mass and metals to the existing CGM.

\section{Summary and conclusions}\label{s:concs}

This work is part of a long-term project to map and analyze AGN-driven winds at the specific evolutionary stage of post starburst E+A galaxies. In this paper, we present new observations of SDSS J003443.68+251020.9, at z=0.118. The new KCWI spatially-resolved spectroscopy presented here allow us to study the gas properties of the system throughout the entire field of view. Our results can be summarized as follows:
\begin{enumerate}
\item The system consists of two galaxies. The primary galaxy shows prominent Balmer absorption lines and no contribution from O and B-type stars, suggestive of a post starburst E+A galaxy.
\item We model the star formation history of the primary galaxy and find a starburst that started 400 Myrs ago, with a peak SFR of about 120 $\mathrm{M_{\odot}}$/yr, with an exponential decrease to less than 10\% of its initial value after 200 Myrs. Currently, the system is fully quenched and there is no ongoing SF. The system shows a bulge-dominated morphology and we estimate its stellar mass to be roughly $10^{10.8}\,\mathrm{M_{\odot}}$.
\item We detect two kinematic components within the primary galaxy, both of which are ionized by the active BH. The narrow component is classified as a LINER, and the broader component as a Seyfert. The narrow component is fully consistent with a typical NLR in type II AGN. The velocity dispersion of the broader component (FWHM$\sim$900 km/sec) exceeds the escape velocity of the galaxy, which suggests a galactic-scale outflow.
\item The KCWI observations reveal gas that extends to distances of 17 kpc from the central galaxy, far beyond the regions in which stars are detected ($\sim$3 kpc), with a conic-shaped geometry. This gas is ionized by the central BH, and its dynamics suggest an AGN-driven outflow.
\item We construct a self-consistent photoionization model for the three gas components (NLR, high-velocity central component, and extended gas). The central source has a bolometric luminosity of about $10^{45}$ erg/sec and a hard X-ray continuum. We estimate a BH mass of $10^{8.1}\,\mathrm{M_{\odot}}$, accreting at a few percent of the Eddington luminosity. 
\item We construct a dynamical model for the gas in the gas cone, and correct for projection effects. The velocity of the gas outside the galaxy is $v(r) = 800 \times \Big( \frac{r}{1\,\mathrm{kpc}} \Big)^{-0.5}$ km/sec.
\item The combination of photoionization and dynamical models for the gas in the cone explains all the observed line luminosities and line profiles. Together they show that the outflow forms a continuous flow, with a constant mass outflow rate of approximately 24 $\mathrm{M_{\odot}}$/yr. The continuity of the flow allows us to put a limit on the activity time of the central BH, which we find to be $\sim$60 Myrs.
\item We are able to measure the gas mass in all the components. The mass outside the galaxy, roughly $10^{9}\,\mathrm{M_{\odot}}$, is an order of magnitude higher than the gas mass within the galaxy (NLR, central high velocity component, neutral component traced by dust). 
\item The total kinetic power of the wind, measured for the central high velocity component and the gas in the cone, is roughly 0.01$L_{\mathrm{bol}}$.
\end{enumerate}

Our new results suggest that we are witnessing a short-lived phase, in which the AGN is successfully removing most of the gas that was in the primary galaxy. Although the current AGN episode cannot be used to determine whether the AGN have contributed to the abrupt quenching of SF in the galaxy, we find that in a single episode, the AGN is able to remove most of the gas from its host galaxy. This gas continues to expand outwards and will probably mix with the surrounding CGM after a few hundrends Myrs, enriching it with the metals it carried out of the galaxy. 

Post starburst E+A galaxies offer an advantage over other galaxy samples for studying AGN-driven outflows. Since these systems are fully quenched, with no ongoing SF, one can study the observed winds in the context of pure AGN feedback, with no contributions from supernovae-driven winds. Their narrow stellar age distribution, which is dominated by the short lifetime of A-type stars, allows us to put such systems on a single timeline, where time is measured as the onset (or termination) of the recent starburst, and compare various wind properties as a function of time. We are currently involved in a detailed analysis of additional post starburst galaxies showing evidence of massive AGN-driven winds, and results will be reported in a forthcoming publication.

\section*{Acknowledgments}
We thank B. Trakhtenbrot and D. Poznanski for useful discussions regarding the manuscript.
Funding for this work was provided by the Israel Science Foundation grant 284/13. 
SC gratefully acknowledges support from Swiss National Science Foundation grant PP00P2-163824.
The spectroscopic analysis was made using IPython \citep{perez07}. We also used the following Python package: astropy\footnote{www.astropy.org/}.

This work made use of SDSS-III\footnote{www.sdss3.org} data. Funding for SDSS-III has been provided by the Alfred P. Sloan Foundation, the Participating Institutions, the National Science Foundation, and the U.S. Department of Energy Office of Science. SDSS-III is managed by the Astrophysical Research Consortium for the Participating Institutions of the SDSS-III Collaboration including the University of Arizona, the Brazilian Participation Group, Brookhaven National Laboratory, Carnegie Mellon University, University of Florida, the French Participation Group, the German Participation Group, Harvard University, the Instituto de Astrofisica de Canarias, the Michigan State/Notre Dame/JINA Participation Group, Johns Hopkins University, Lawrence Berkeley National Laboratory, Max Planck Institute for Astrophysics, Max Planck Institute for Extraterrestrial Physics, New Mexico State University, New York University, Ohio State University, Pennsylvania State University, University of Portsmouth, Princeton University, the Spanish Participation Group, University of Tokyo, University of Utah, Vanderbilt University, University of Virginia, University of Washington, and Yale University. 

\bibliographystyle{mn2e}
\bibliography{ref_ea_gal_kcwi}

\clearpage

\appendix
\section{Spectral Fitting}\label{a:app}

\subsection{Global spectral fitting}

\begin{figure*}
\includegraphics[width=0.8\textwidth]{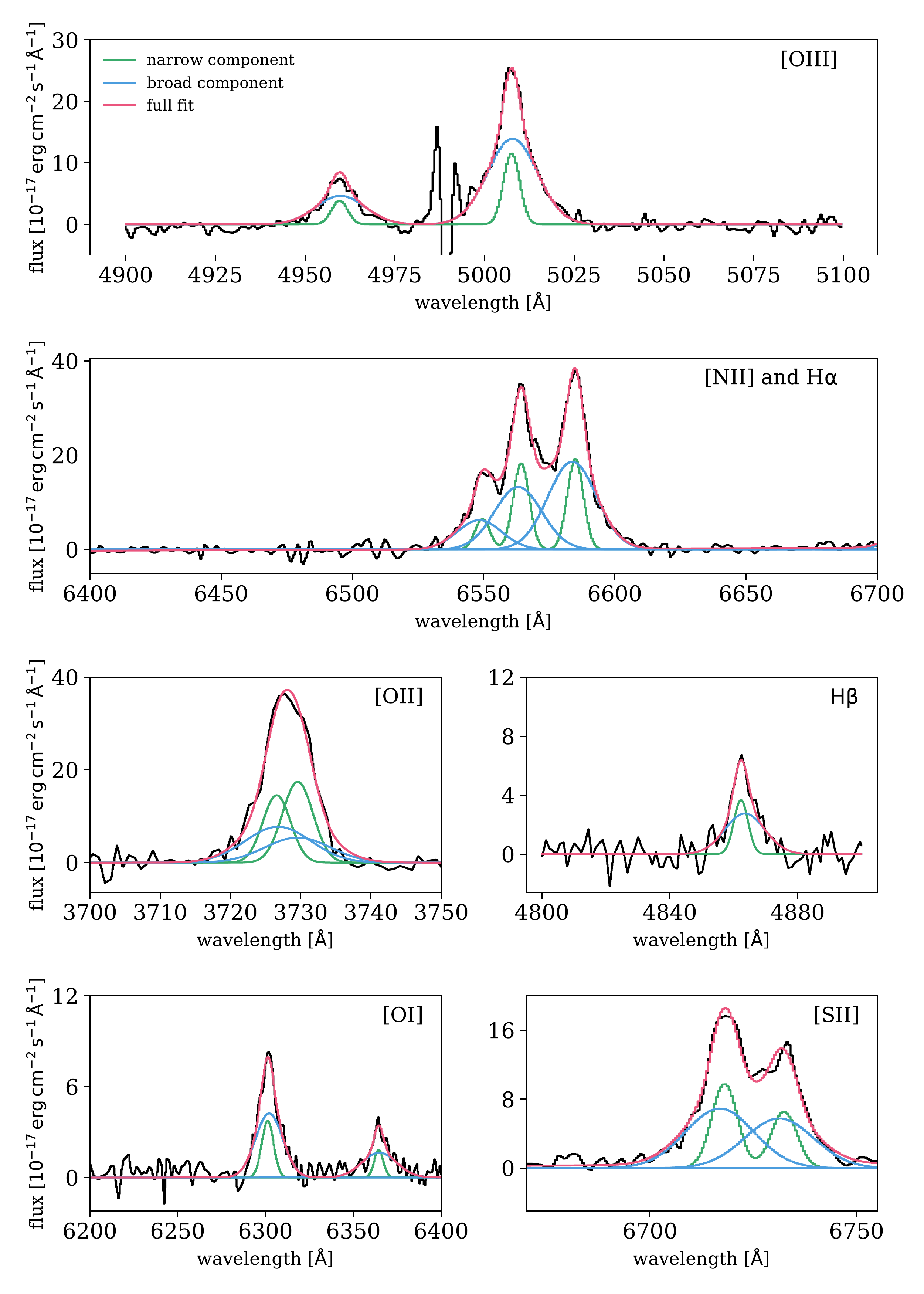}
\caption{The emission-line spectrum (black) of the gas in the galaxy as obtained by subtracting the best-fitting population synthesis model from the global (SDSS) spectrum. We show [OIII], H$\alpha$ and [NII], [OII], H$\beta$, [OI], and [SII] as indicated on each panel. We use two Gaussians to account for the narrow and broad components, and plot these in green (narrow lines) and blue (broad lines). The full fit is plotted in pink.}\label{f:1d_spec_line_fit}
\end{figure*}

We refer to the SDSS spectrum of the primary galaxy as the global spectrum. We subtract the best-fitting stellar model from the global spectrum and obtain the emission-line spectrum. We show six parts of this spectrum in figure \ref{f:1d_spec_line_fit}: [OIII]~$\lambda \lambda$ 4959,5007\AA, $\mathrm{H\alpha}$~$\lambda$ 6563\AA\, and [NII]~$\lambda \lambda$ 6548,6584\AA, [OII]~$\lambda \lambda$ 3725,3727\AA\, and $\mathrm{H\beta}$~$\lambda$ 4861\AA, and [OI]~$\lambda \lambda$ 6300,6363\AA\, and [SII]~$\lambda \lambda$ 6717,6731\AA\, (hereafter $\mathrm{[OIII]}$, $\mathrm{H\alpha}$, $\mathrm{[NII]}$, $\mathrm{[OII]}$, $\mathrm{H\beta}$, $\mathrm{[OI]}$, and $\mathrm{[SII]}$). 
The Balmer lines and the forbidden line profiles show narrow as well as broad components.

We model each emission line as a sum of two Gaussians - one which represents the narrow component and one that represents the broader component. We perform a joint fit to all the emission lines under several constrains: (1) we force the intensity ratio of the emission lines [OIII]~$\lambda \lambda$ 4959,5007\AA\, and [NII]~$\lambda \lambda$ 6548,6584\AA\, to the theoretical ratio of 3:1, (2) we tie the central wavelengths of all the narrow lines so that the gas has the same systematic velocity, we do the same for the broader lines, and (3) we force the widths of all the narrow lines to show the same velocity dispersion, and we do the same for the broader lines. We show in figure \ref{f:1d_spec_line_fit} the best-fitting profiles, where we mark the narrow lines with green, the broader lines with blue, and the full fit with pink. 

Using the best fitting model, we derive the kinematic properties of the emitting gas. The FWHM velocity of the narrow lines is 320 km/sec, and the width of the broad lines is 980 km/sec. The narrow and broad components show roughly the same systematic velocity, and both are blueshifted by about 90 km/sec from the systematic redshift of the galaxy which we measure from the stellar Balmer absorption lines. Following the same arguments and analysis used in \citet[see section 3.3.2 there]{baron17b}, we find that the velocity dispersion of the broader component exceeds the escape velocity from the galaxy, even under the assumption that the gas mass is half of the stellar mass in that system. 

Table \ref{t:meas} gives line intensities, reddening corrected line luminosities, and derived fit parameters for all measured emission lines. The uncertainties of the widths and central wavelengths are obtained from the $\chi^{2}$ minimisation process, and are tied to each other as discussed above. We propagate these uncertainties and the uncertainties on the dust reddening to calculate the uncertainties in line intensity and extinction corrected luminosity.

\begin{table}
	\centering 
	\tiny
	\tablewidth{0.8\linewidth} 
\begin{tabular}{|p{2.8cm}||c|r|}
 Emission & Intensity & Luminosity \\
 Line  &  [$10^{-17}\,\mathrm{erg\,s^{-1}\,cm^{-2}}$]    &  [$\mathrm{erg\,s^{-1}}$]\\
 \hline
 \hline
 \multicolumn{3}{|l|}{Narrow lines: $\mathrm{\sigma=135 \pm 20\,km\,s^{-1}}$, $\mathrm{WL([OIII]\lambda5007)=5007.7\pm 0.2}$\AA} \\
 \hline
$\mathrm{H\beta}$				& $30.7 \pm 2.9$   & $4.96 \pm 0.51 \times 10^{40}$ \\
$\mathrm{[OIII]\lambda 4959}$   & $21.6 \pm 2.3$   & $3.62 \pm 0.38 \times 10^{40}$ \\
$\mathrm{[OIII]\lambda 5007}$   & $65.4 \pm 5.0$   & $10.75 \pm 0.82 \times 10^{40}$ \\
$\mathrm{[OI]\lambda 6300}$     & $31.4 \pm 5.6$   & $3.39 \pm 0.61 \times 10^{40}$ \\
$\mathrm{[NII]\lambda 6548}$    & $45.7 \pm 3.5$   & $4.63 \pm 0.35 \times 10^{40}$ \\
$\mathrm{H\alpha}$              & $140.0 \pm 11.1$   & $14.2 \pm 1.1 \times 10^{40}$ \\
$\mathrm{[NII]\lambda 6584}$    & $147.4 \pm 10.8$    & $14.8 \pm 1.1 \times 10^{40}$ \\
$\mathrm{[SII]\lambda 6717}$    & $110.8 \pm 7.9$   & $10.81 \pm 0.95 \times 10^{40}$ \\
$\mathrm{[SII]\lambda 6731}$    & $80.1 \pm 6.34$   & $7.79 \pm 0.88 \times 10^{40}$ \\
								&					&  \\
 \hline
 \hline
 \multicolumn{3}{|l|}{Blueshifted broad lines: $\mathrm{\sigma=416\pm 70 \,km\,s^{-1}}$, $\mathrm{WL([OIII]\lambda5007)=5008.0 \pm 1.4}$\AA} \\
 \hline
$\mathrm{H\beta}$				& $19.1 \pm 3.1$   & $21.2 \pm 3.4 \times 10^{41}$ \\
$\mathrm{[OIII]\lambda 4959}$   & $79.9 \pm 5.6$   & $71.8 \pm 5.3 \times 10^{41}$ \\
$\mathrm{[OIII]\lambda 5007}$   & $242.1 \pm 9.7$  & $213.2 \pm 8.5 \times 10^{41}$ \\
$\mathrm{[OI]\lambda 6300}$     & $82 \pm 12$      & $15.6 \pm 2.2 \times 10^{41}$ \\
$\mathrm{[NII]\lambda 6548}$    & $143.3 \pm 5.1$  & $21.2 \pm 0.77 \times 10^{41}$ \\
$\mathrm{H\alpha}$              & $298 \pm 13$     & $60.0 \pm 2.1 \times 10^{41}$ \\
$\mathrm{[NII]\lambda 6584}$    & $430 \pm 15$     & $63.7 \pm 2.2 \times 10^{41}$ \\
$\mathrm{[SII]\lambda 6717}$    & $125.4 \pm 6.4$  & $16.6 \pm 1.3 \times 10^{41}$ \\
$\mathrm{[SII]\lambda 6731}$    & $91.3 \pm 5.8$   & $12.0 \pm 1.2 \times 10^{41}$ \\
								&				   &  \\
 \hline
\end{tabular}
\caption{Best fitting model parameters for the two Gaussian fit. The best fitting intensities are normalized with respect to the median value of the entire spectrum. The central wavelengths and widths of the Gaussians are tied together for a given velocity component. The luminosities are corrected for reddening by foreground ISM-type dust assuming $\mathrm{H\alpha/H\beta}$=2.85.}
\label{t:meas}
\end{table}

\subsection{Spatially-resolved spectral fitting}

\begin{figure*}
\includegraphics[width=0.8\textwidth]{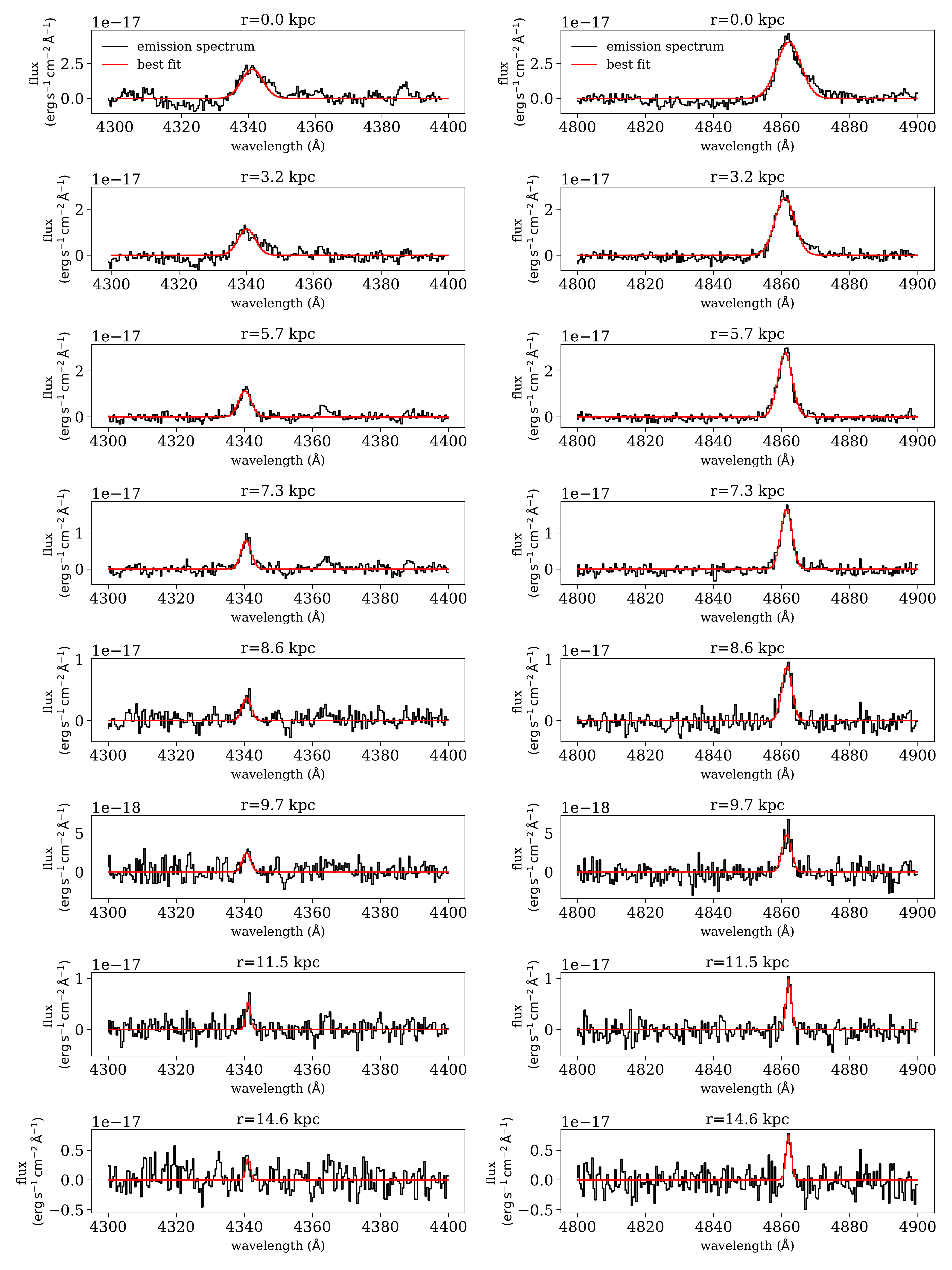}
\caption{Stacked spectra as a function of distance from the primary galaxy, where the left panels show the H$\gamma$ region and the right panels show the H$\beta$ region (black). The best-fitting profiles are marked with red.}\label{f:hbeta_hgamma_fit}
\end{figure*}

In order to increase the SNR of the spatially-resolved spectra, we divide the spaxels into bins with similar distance from the primary galaxy using shells, as described in section \ref{s:gas_properties}. We show in figure figure \ref{f:hbeta_hgamma_fit} the H$\beta$ and H$\gamma$ emission line regions of the binned spectra, as a function of distance from the primary galaxy. In order to reduce the uncertainty in fitting the H$\gamma$ line, we fit simultaneously the H$\beta$ and H$\gamma$ lines with a single Gaussian for each, tying their central wavelengths and velocity dispersions. The best-fitting profiles are marked with red in figure \ref{f:hbeta_hgamma_fit}.

\begin{figure*}
\includegraphics[width=0.9\textwidth]{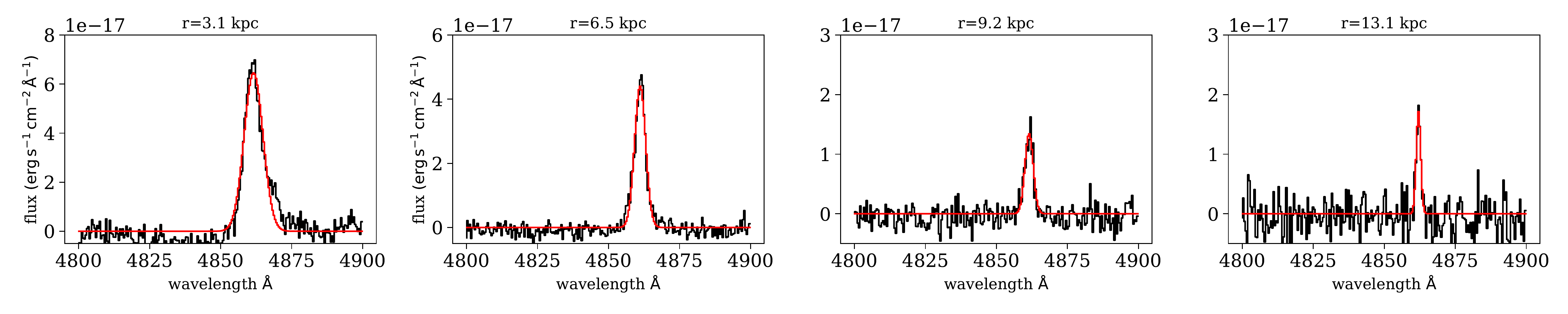}
\includegraphics[width=0.9\textwidth]{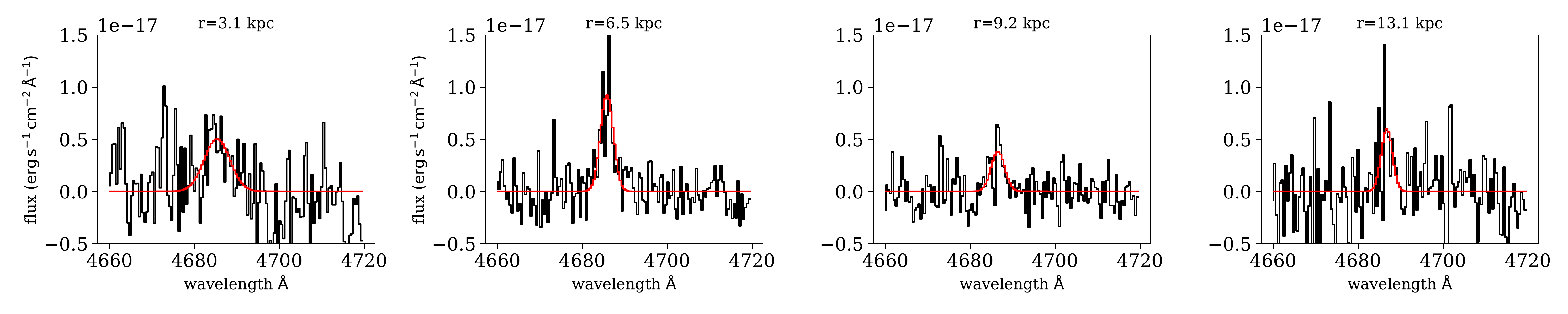}
\caption{Stacked spectra along the gas cone, as a function of distance from the primary galaxy. The upper panels show the H$\beta$ emission line region and the lower panels show the HeII emission line region. The best-fitting profiles are marked with red.}\label{f:hbeta_helium_fit}
\end{figure*}

Since the HeII emission line is even weaker than the H$\gamma$, we bin the spectra into 4, rather than 8, bins, as a function of distance from the primary galaxy. For each bin, we fit simultaneously the H$\beta$ and the HeII lines, which we model with a single Gaussian, and we tie their central wavelengths and velocity dispersions. We show the stacked spectra, centered around the H$\beta$ emission line in the top panels of figure \ref{f:hbeta_helium_fit}, and around the HeII emission line in the bottom panels of the figure. The best-fitting profiles are marked with red. 

We list in table \ref{t:flux_meas} the luminosities of the lines measured in eight shells and the dust reddening calculated from the H$\beta$/H$\gamma$ ratio. In table \ref{t:flux_meas_heii} we list the HeII and H$\beta$ luminosities measured in four shells. Finally, table \ref{t:w80_meas} lists the W80 values measured for the [OIII] emission line as a function of distance from the central galaxy.

\begin{table}
	\centering 
	\tiny
	\tablewidth{0.8\linewidth} 
\begin{tabular}{|l|l|l|l|l|}
 Distance & [OIII] luminosity &        H$\beta$ luminosity &      H$\gamma$ luminosity &   $\mathrm{E}(B-V)$ \\
 kpc &      $10^{40}$ erg/sec & $10^{39}$ erg/sec & $10^{39}$ erg/sec & mag \\
 \hline
 \hline

$0.0 \pm 0.38$ & $4.63 \pm 0.58$ & $9.2 \pm 1.3$ & $3.33 \pm 0.57$ & $0.531 \pm 0.066$ \\
$3.1 \pm 1.5$ & $3.35 \pm 0.25$ & $4.63 \pm 0.40$ & $1.75 \pm 0.17$ & $0.438 \pm 0.076$ \\
$5.72 \pm 0.94$ & $3.34 \pm 0.11$ & $3.76 \pm 0.14$ & $1.541 \pm 0.072$ & $0.281 \pm 0.085$ \\
$7.34 \pm 0.54$ & $1.954 \pm 0.093$ & $1.798 \pm 0.095$ & $0.753 \pm 0.050$ & $0.234 \pm 0.081$ \\
$8.62 \pm 0.62$ & $1.021 \pm 0.059$ & $0.841 \pm 0.056$ & $0.357 \pm 0.030$ & $0.195 \pm 0.079$ \\
$9.72 \pm 0.53$ & $0.566 \pm 0.076$ & $0.419 \pm 0.064$ & $0.194 \pm 0.035$ & $0.05 \pm 0.10$ \\
$11.5 \pm 1.13$ & $0.53 \pm 0.10$ & $0.48 \pm 0.10$ & $0.228 \pm 0.064$ & $0.00 \pm 0.10$ \\
$14.6 \pm 1.75$ & $0.37 \pm 0.14$ & $0.38 \pm 0.16$ & $0.178 \pm 0.098$ & $0.00 \pm 0.13$ \\

 \hline
\end{tabular}
\caption{Spatially-resolved emission line measurements for the eight different shells. The listed luminosities are not reddening corrected since this depends on the assumed dust geometry (see section \ref{s:photo_gas_cone}).}
\label{t:flux_meas}
\end{table}

\begin{table}
	\centering 
	\tiny
	\tablewidth{0.8\linewidth} 
\begin{tabular}{|l|l|l|}
 Distance [kpc] & H$\beta$ luminosity [$10^{39}$ erg/sec] & HeII luminosity [$10^{39}$ erg/sec]\\
 \hline
 \hline 
 $2.75 \pm 1.50$ & $7.13 \pm 0.81$ & $1.74 \pm 0.49$ \\
 $6.53 \pm 1.74$ & $7.02 \pm 0.92$ & $1.17 \pm 0.45$ \\
 $9.17 \pm 1.20$ & $1.88 \pm 0.75$ & $0.59 \pm 0.23$ \\
 $13.0 \pm 3.51$ & $1.31 \pm 0.41$ & $0.77 \pm 0.29$ \\
 \hline
\end{tabular}
\caption{Spatially-resolved emission line measurements for the four different shells. The luminosities we list are not corrected for dust, since they depend on the assumed dust geometry (see section \ref{s:photo_gas_cone}).}
\label{t:flux_meas_heii}
\end{table}

\begin{table}
	\centering 
	\tablewidth{0.8\linewidth} 
\begin{tabular}{|l|l|}
 Distance [kpc] & W80 [km/sec]\\
 \hline
 \hline
$ 0.00^{+0.78}_{-0} $ & $ 1040^{+215}_{-130} $ \\ 
 \hline
$ 1.57^{+0.65}_{-0.78} $ & $ 1013^{+245}_{-133} $ \\
 \hline
$ 2.87^{+0.27}_{-0.65} $ & $ 720^{+191}_{-92} $ \\
 \hline
$ 3.42^{+0.51}_{-0.27} $ & $ 538^{+147}_{-75} $ \\
 \hline
$ 4.45^{+0.69}_{-0.51} $ & $ 292^{+41}_{-23} $ \\
 \hline
$ 5.83^{+0.51}_{-0.69} $ & $ 240^{+36}_{-20} $ \\
 \hline
$ 6.87^{+0.38}_{-0.51} $ & $ 219^{+24}_{-15} $ \\
 \hline
$ 7.64^{+0.45}_{-0.38} $ & $ 208^{+28}_{-15} $ \\
 \hline
$ 8.56^{+0.31}_{-0.45} $ & $ 206^{+24}_{-14} $ \\
 \hline
$ 9.19^{+0.39}_{-0.31} $ & $ 192^{+31}_{-17} $ \\
 \hline
$ 9.98^{+0.76}_{-0.39} $ & $ 185^{+34}_{-15} $ \\
 \hline
$ 11.5^{+1.5}_{-0.76} $ & $ 172^{+25}_{-11} $ \\
 \hline
$ 14.5^{+1.5}_{-1.5} $ & $ 110^{+16}_{-11} $ \\

 \hline
\end{tabular}
\caption{$W_{80}(\mathrm{[OIII])}$ as a function of distance from the center of the primary galaxy. $W_{80}$ is defined as the width of the [OIII] that contains 80 percent of its integrated flux: $W_{80} = v_{90} - v_{10}$, where $v_{10}$ and $v_{90}$ are the velocities that correspond to the 10th and 90th percentiles of the integrated [OIII] flux. The uncertainties on the distance represent the different shells we used, and the uncertainties on W80 represent the 85th and 75th percentiles of the flux. }
\label{t:w80_meas}
\end{table}

\end{document}